\documentclass{aa}  

\usepackage{graphicx}
\usepackage{afterpage}
\usepackage{booktabs}
\usepackage{subcaption}
\usepackage{longtable}
\usepackage{xcolor}
\usepackage[switch]{lineno}
\usepackage{float}
\usepackage{makecell}
\usepackage{placeins}
\usepackage{natbib}
\usepackage{txfonts}
\usepackage[colorlinks=true, allcolors=blue]{hyperref}

\usepackage{etoolbox}

\begin{document} 

   \title{Exploring the most extreme gamma-ray blazars using broadband spectral energy distributions}

   \author{M. L\'ainez
          \inst{1} \thanks{\email{malainez@ucm.es}}
          \and
          M. Nievas-Rosillo\inst{2,3}
          \thanks{ \email{mireia.nievas@iac.es}}
          \and
          A. Dom\'inguez\inst{1} \thanks{\email{alberto.d@ucm.es}}
          \and
          J. L. Contreras\inst{1}
          \and
          J. Becerra Gonz\'alez\inst{2,3}
          \and
          A. Dinesh\inst{1}
          \and
          V. S. Paliya\inst{4}
          }

   \institute{IPARCOS and Department of EMFTEL, Universidad Complutense de Madrid, E-28040 Madrid, Spain
         \and
             Instituto de Astrof\'isica de Canarias, E-38205 La Laguna, Tenerife, Spain
        \and
            Universidad de La Laguna, Dept. Astrof\'isica, E-38206 La Laguna, Tenerife, Spain
        \and
            Inter-University Centre for Astronomy and Astrophysics (IUCAA), SPPU Campus, 411007, Pune, India
             }

   \date{Received 17 February 2025 / Accepted 10 July 2025}

  \abstract
   {Extreme high-synchrotron peaked blazars (EHSPs) are rare high-energy sources characterised by synchrotron peaks beyond 10$^{17}$ Hz in their spectral energy distributions (SEDs). Their extreme properties challenge conventional blazar emission models and provide a unique opportunity to test the limits of particle acceleration and emission mechanisms in relativistic jets. However, the number of identified EHSPs is still small, limiting comprehensive studies of their population and characteristics.}
   {This study aims to identify new EHSP candidates and characterise their emission properties, including synchrotron peak frequencies, Compton dominance, and jet environments. It also examines how EHSPs fit within the broader framework of the blazar sequence, providing insights into their role in the population of active galactic nuclei.} 
   {A sample of 124 $\gamma$-ray blazars was analysed, selected for their high synchrotron peak frequencies and $\gamma$-ray emission properties, with a focus on sources showing low variability and good broadband data coverage. Their SEDs were constructed using archival multi-wavelength data from the SSDC SED Builder service, supplemented with recent \textit{Swift}-UVOT, \textit{Swift}-XRT, and \textit{Fermi}-LAT observations. The SEDs were modelled with a one-zone synchrotron/synchrotron-self-Compton framework, classifying sources by synchrotron peak frequency. EHSP properties are compared to other blazar populations, and their detectability with the Cherenkov Telescope Array Observatory (CTAO) is assessed.}
   {We identify 66 new EHSP candidates, significantly expanding the known population. A clear correlation between synchrotron peak frequency and the magnetic-to-kinetic energy density ratio is found, with the most extreme EHSPs nearing equipartition. This indicates that as the synchrotron peak shifts to higher frequencies, the energy stored in the magnetic field becomes comparable to that of the relativistic electrons, suggesting a more balanced and energetically efficient jet environment in the most extreme blazars. Host galaxy emission is detected in many sources, but no significant differences are observed between elliptical and lenticular hosts. Finally, our analysis suggests that nine high-synchrotron peaked/EHSPs could be observed by CTAO at $>5\sigma$ (20 at $>3\sigma$) in 20-hour exposures, a feasible integration time for Imaging Atmospheric Cherenkov Telescopes, indicating that while the overall detection rate remains modest, a subset of these sources is within reach of next-generation very-high-energy gamma-ray instruments.}
   {}


   \maketitle

\section{Introduction}

Active galactic nuclei (AGNs) are extreme cosmic sources, powered by matter accretion onto a supermassive black hole \citep[SMBH, e.g.][]{2021agnf.book.....C}. Some develop relativistic jets extending to kiloparsec scales \citep[e.g.][]{2010LNP...794..173M}. Unification models classify radio-loud AGNs by jet viewing angle \citep[e.g.][]{1995PASP..107..803U}, with blazars, jets aligned close to our line of sight, dominating the extragalactic $\gamma$-ray sky \citep{Abdollahi_2020}.

Blazars have broadband spectral energy distributions (SEDs) dominated by non-thermal emission from radio to $\gamma$ rays \citep[e.g.][]{1998MNRAS.299..433F, 2017FrASS...4...35P}. Their SEDs feature a double-peaked structure: the lower-energy peak (infrared to X-ray) arises from synchrotron emission by relativistic jet electrons, while the origin of the higher-energy $\gamma$-ray peak remains debated \citep[e.g.][]{2010ApJ...716...30A}, with both hadronic and leptonic processes proposed.

Leptonic models explain the higher-energy peak of blazar SEDs as inverse Compton (IC) scattering, where high-energy electrons interact with low-energy photons \citep[e.g.][]{2020A&A...637A..86M, 2021A&A...655A..89M}. This occurs mainly through synchrotron-self-Compton (SSC), where electrons scatter their own synchrotron photons \citep[e.g.][]{1992ApJ...397L...5M, 2008ApJ...686..181F}, and external Compton, where electrons scatter external photons from sources like the cosmic microwave background, broad-line region (BLR), dusty torus, disc, stars, or different jet layers \citep[e.g.][]{1994ApJ...421..153S, 2019ApJ...874...47V}. Hadronic models instead attribute $\gamma$-ray emission to high-energy proton interactions with ambient matter or radiation fields and to synchrotron radiation from relativistic protons in strong magnetic fields \citep[e.g.][and references therein]{1992A&A...253L..21M, 2013ApJ...768...54B}. Lepto-hadronic models combine both leptonic and hadronic processes \citep[e.g.][]{2003APh....18..593M, 2018ApJ...863...98P}. The main acceleration mechanisms proposed for ultra-relativistic particles in blazar jets include shock acceleration via repeated shock crossings and magnetic turbulence \citep[e.g.][]{1998A&A...333..452K, 2020MNRAS.498..599T}, as well as magnetic reconnection \citep[e.g.][]{2014ApJ...783L..21S} and stochastic acceleration \citep[e.g.][]{1987PhR...154....1B, 2017ApJ...842...39L}.

Blazars are classified based on their optical spectra into flat spectrum radio quasars (FSRQs) and BL Lacertae objects (BL Lacs). While FSRQs exhibit strong, broad emission lines (EW $> 5$ \r{A}), BL Lacs have an almost featureless continuum \citep[e.g.][]{1996MNRAS.281..425M}, indicating that external photon sources like the BLR, dusty torus, or disc are subdominant, with emission mainly from synchrotron/SSC \citep[e.g.][]{2011MNRAS.414.2674G}. In the most extreme BL Lacs, where the synchrotron peak is at higher frequencies, the host galaxy's thermal emission becomes visible in the optical/ultraviolet (UV) range, allowing absorption lines to be detected. Blazars are also categorised by their synchrotron peak frequency \citep[e.g.][]{1995ApJ...444..567P, 2010ApJ...716...30A}, from low-synchrotron peaked (LSP, $\nu_{sync}^{peak} < 10^{14}$ Hz) to intermediate (ISP, $10^{14}$ Hz $\leq \nu_{sync}^{peak} < 10^{15}$ Hz), high (HSP, $10^{15}$ Hz $\leq \nu_{sync}^{peak} < 10^{17}$ Hz), and extreme high-synchrotron peaked (EHSP, $\nu_{sync}^{peak} \geq 10^{17}$ Hz), with peaks often reaching the X-ray band.

Blazars often show variability on timescales from minutes to years, differing across energy bands \citep[e.g.][]{1996ASPC..110..391U, 2008Natur.452..966M}. HSPs and EHSPs vary significantly in X-rays but show little to no large-scale variability in the $\gamma$-ray band, whereas LSPs and ISPs are generally more variable at high-energy (HE, 100 MeV $< E <$ 100 GeV) $\gamma$ rays. This pattern may result from the low $\gamma$-ray luminosity of HSPs and EHSPs, as their apparent lack of variability could stem from the limited sensitivity of \textit{Fermi}-LAT and Imaging Atmospheric Cherenkov Telescopes (IACTs) to low-flux sources \citep{2011ApJ...743..171A}.

The most luminous blazars, FSRQs, are mainly LSPs, while higher synchrotron peak frequency blazars are mostly lower-luminosity BL Lacs \citep[e.g.][]{2012MNRAS.420.2899G}. This anti-correlation between luminosity and synchrotron peak frequency defines the `blazar sequence' \citep{1998MNRAS.299..433F, 2017MNRAS.469..255G}, though observational biases and sample selection may challenge its validity \citep[e.g.][]{Padovani_2007, 2012MNRAS.420.2899G, 2022MNRAS.511.4697P}.

Characterised by radiatively inefficient accretion discs, EHSPs occupy the end of the blazar sequence \citep[e.g.][]{2011MNRAS.414.2674G}. Their extreme synchrotron peak frequencies challenge standard models, requiring extreme conditions for the ultra-relativistic particles responsible for their emission \citep[e.g.][]{2020NatAs...4..124B}. Studying EHSPs enhances our understanding of jet acceleration and non-thermal emission. Their high energies \citep[up to tens of tera-electronvolts,][]{2020NatAs...4..124B} make them valuable for cosmological studies, including the extragalactic background light \citep[EBL, e.g.][]{2015ApJ...813L..34D, 2019ApJ...885..137D,2024ApJ...975L..18G}, blazar evolution \citep[e.g.][]{2014ApJ...780...73A}, the extragalactic $\gamma$-ray background \citep[e.g.][]{2015ApJ...800L..27A, 2019ApJ...882L...3P}, and the intergalactic magnetic field \citep[e.g.][]{2015MNRAS.451..611B}. As potential very-high-energy (VHE, $E > 100$ GeV) emitters, they are prime targets for IACTs. Although blazars, particularly BL Lacs, dominate the extragalactic $\gamma$-ray sky, only a few dozen EHSPs are known \citep{NIEVAS2022, 2020ApJS..247...16A}.

This work examines 124 blazars selected as possible EHSPs by modelling their broadband SEDs with a one-zone synchrotron/SSC model to classify them by synchrotron peak frequency and assess their physical properties within the broader blazar population.

The paper is organised as follows: Section~\ref{sec:sample} describes source selection and multi-wavelength data, Section~\ref{sec:modeling} details SED modelling, and Section~\ref{sec:results} discusses EHSP properties. Section~\ref{sec:cta-significance} presents detectability predictions with the Cherenkov Telescope Array Observatory (CTAO), and Section~\ref{sec:summary} provides a summary and conclusions.

Throughout this paper, we adopt the following cosmological parameters: Hubble constant $H_0 = 67.8$ km s$^{-1}$ Mpc$^{-1}$, matter density parameter $\Omega_{M,0} = 0.307$, baryon density parameter $\Omega_{b,0} = 0.0483$, cosmic microwave background temperature $T_{\text{CMB},0} = 2.725$ K, effective number of relativistic degrees of freedom $N_{\text{eff}} = 3.05$, and a total neutrino mass sum of 0.06 eV.

\section{Sample selection and multi-wavelength data}
\label{sec:sample}

\subsection{Source selection}

We selected blazars from the Second Brazil-ICRANet Gamma-ray Blazars catalogue \citep[2BIGB catalogue;][]{2020MNRAS.493.2438A}, which contains 1160 $\gamma$-ray blazars along with their $\gamma$-ray SEDs. These SEDs were obtained from the first 11 years of observations by the \textit{Large Area Telescope} (LAT) aboard the \textit{Fermi Gamma-ray Space Telescope}, a space-based instrument designed to detect $\gamma$ rays in the energy range from approximately 20 MeV to over 300 GeV \citep{2009ApJ...697.1071A}. The 2BIGB catalogue was produced through a $\gamma$-ray likelihood analysis of all sources in the 3HSP catalogue \citep{2019A&A...632A..77C}, which remains the largest collection of HSPs, EHSPs, and blazar candidates of uncertain type (BCUs; that is, sources detected by \textit{Fermi}-LAT that show characteristics typical of blazars, as a double peaked SED, but lack definitive classification as either BL Lac or FSRQ).

To avoid contamination of diffuse emission from the Galactic plane, we selected only blazars located at high latitudes ($|b|>10^\circ$). We also required that all sources in our sample have a redshift estimate, even if photometric, along with flux measurements in the optical/UV, X-ray, and $\gamma$-ray bands. With these criteria, we obtained a preliminary sample of 657 $\gamma$-ray blazars. We cross-referenced these sources with the \textit{Fermi}-LAT Fourth Source catalogue Data Release 4 \citep[4FGL-DR4,][]{2023arXiv230712546B}, which lists $\gamma$-ray sources detected by \textit{Fermi}-LAT over its first 14 years.

\begin{figure*}
	\includegraphics[width=\textwidth]{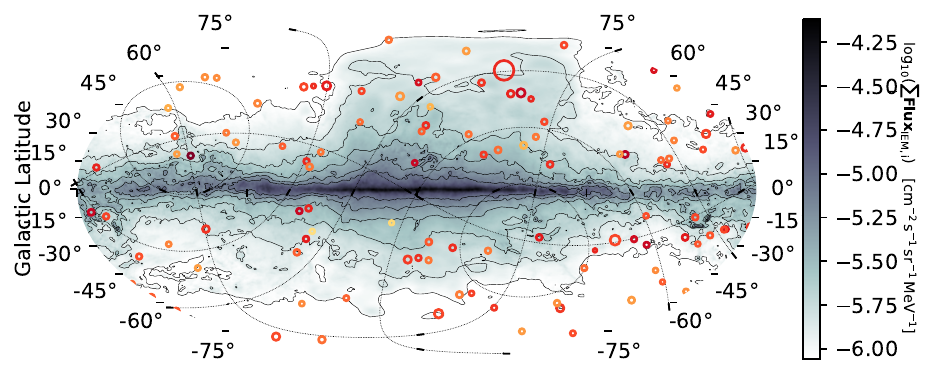}
    \caption{Sky location of 124 sources selected for our analysis in Galactic coordinates. Each circle represents a source, with its colour corresponding to the flux (in log scale, with red tones for brighter sources). The size of the circle is proportional to $1/\sqrt{z}$. For reference, the diffuse $\gamma$-ray emission from the Galaxy is shown in blue tones.}
    \label{fig:skymap}
\end{figure*}

Since we used archival data from multiple telescopes to reconstruct the broadband SEDs of the selected blazars, and these data are not always taken simultaneously, additional cuts were applied to select sources with low variability across different energy bands. The variability index from the 4FGL-DR4 catalogue was used to identify sources with low variability in the \textit{Fermi}-LAT energy range. This index is calculated for each source as twice the sum of the log-likelihood differences between the flux in various time intervals and the average flux over the entire catalogue interval \citep{2020ApJ...892..105A}. Only sources with a variability index below 27.69, the 99\% confidence threshold in 4FGL-DR4, were included.

To assess the variability of the sources in the X-ray band, data from the X-ray Telescope aboard the Neil Gehrels \textit{Swift} Observatory \citep[\textit{Swift}-XRT,][]{2005SSRv..120..165B} were analysed to construct light curves. A Bayesian block analysis \citep{2013ApJ...764..167S}, as implemented in \texttt{astropy} \citep{2022ApJ...935..167A, 2018AJ....156..123A, 2013A&A...558A..33A}, was used to distinguish sources with a single block (likely non-variable) from those with multiple blocks (variable). We adopted a false probability $p_0=0.01$ and a systematic uncertainty of 10\%. Only sources with a single Bayesian block in their light curve were included, resulting in 124 selected blazars. However, gaps in the temporal sampling may affect variability estimates. In general, variability studies benefit from frequent and continuous monitoring, which allows for better characterisation of the synchrotron component. However, for EHSPs, the optical/UV bands often contain a significant contribution from the host galaxy’s thermal emission \citep[e.g.][]{2000ApJ...532..740S}, reducing the sensitivity to AGN variability. As a result, no variability-based selection criteria were applied in this energy range. A sky-map showing the distribution of the 124 selected blazars is presented in Figure~\ref{fig:skymap}.

\begin{figure*}
	\includegraphics[width=\textwidth]{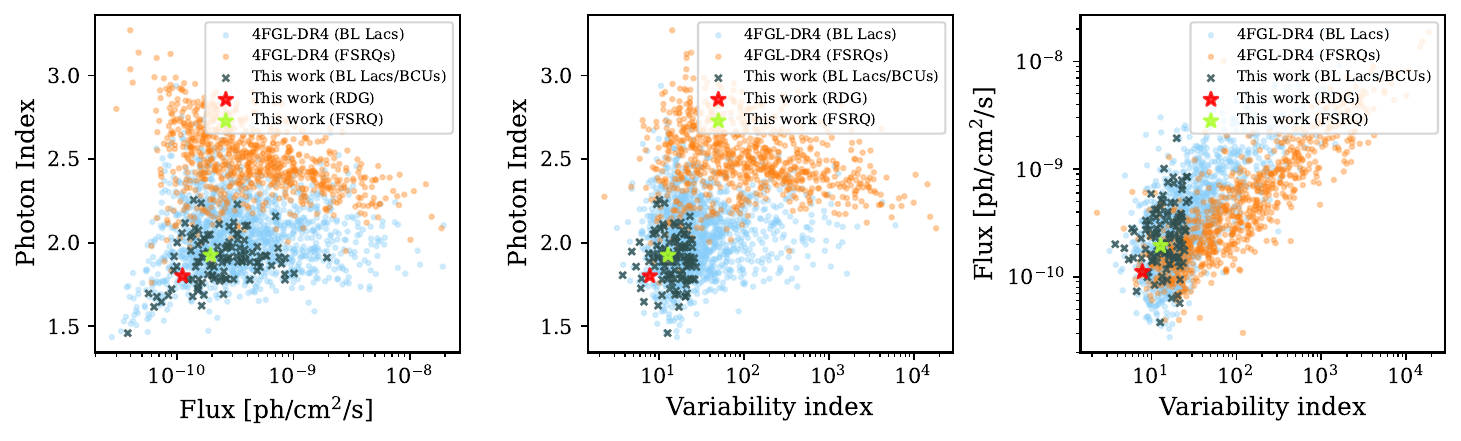}
    \caption{Comparison of the 124 sources in our sample with all BL Lacs and FSRQs in the 4FGL-DR4 catalogue. Left: Photon spectral index versus the integral photon flux from 1 to 100 GeV. Middle: Photon spectral index versus the variability index. Right: Variability index versus the integral photon flux from 1 to 100 GeV. The photon index, flux, and variability index values are from the 4FGL-DR4 catalogue. We represent all the BL Lac sources from the 4FGL-DR4 catalogue with blue circles and the FSRQs with orange circles. The sources selected for this work are marked with dark crosses, the radiogalaxy of the sample (4FGL J1518.6+0614) is marked with a red star, and the FSRQ (4FGL J0132.7$-$0804) is marked with a green star.}
    \label{fig:4lac}
\end{figure*}

We did not apply any additional requirements regarding the class of the sources in 4FGL. Of the 124 sources in the final sample, 93 are classified as BL Lacs in the 4FGL-DR4 catalogue \citep{2023arXiv230712546B}, one as an FSRQ (4FGL J0132.7$-$0804, although it is likely misclassified; see Section \ref{sec:sed_modeling_results}), one as a radiogalaxy (4FGL J1518.6+0614), and 29 as BCUs. Most sources in the sample are BL Lacs since the blazars listed in the 3HSP catalogue show characteristics commonly associated with HSP and EHSP blazars. Figure~\ref{fig:4lac} shows the dependence of the spectral index on flux and the variability index, as well as the relationship between the variability index and flux, for all sources in our sample and all sources classified as BL Lacs or FSRQs in the 4FGL catalogue. As expected, the sources selected for our study tend to have harder spectra than is typically found in FSRQs, as most of them belong to the BL Lac class.

\begin{figure}
    \centering
	\includegraphics[width=\columnwidth]{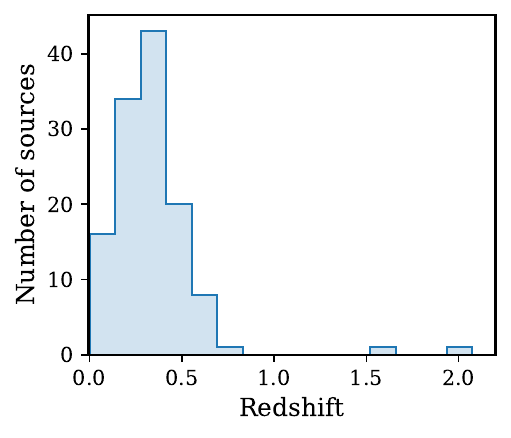}
    \caption{Redshift distribution of the 124 sources of the sample.}
    \label{fig:redshift}
\end{figure}

The redshift distribution of the blazar sample is shown in Figure~\ref{fig:redshift}. They come from the 2BIGB \citep{2020MNRAS.493.2438A} and 4LAC-DR3 \citep{2020ApJ...892..105A, 2022ApJS..263...24A} catalogues, as well as from \citet{goldoni_2021_5512660} and \citet{2021ApJS..253...46P}. Of these, 44 (35.5\%) are from 4LAC, which does not specify whether they are spectroscopic or photometric, while 38 (30.6\%) from 2BIGB are all photometric. The remaining redshifts, 33 (26.6\%) from \citet{2021ApJS..253...46P} and nine (7.3\%) from \citet{goldoni_2021_5512660}, are spectroscopic. Most of the sources in our sample ($\sim 86\%$) have a redshift of $z < 0.5$, although some sources reach much higher redshifts, with the farthest blazar in the sample located at $z = 2.075$ (4FGL J0323.7$-$0111, associated with 1RXS J032342.6$-$011131), though this redshift, given in 4LAC, may be uncertain as the optical spectrum of the associated source is continuum-dominated.

\subsection{Multi-wavelength data}
\label{sec:data}

For the construction of the broadband SEDs of all sources in the sample, we gathered data from various telescopes covering different energy bands. In the X-ray band, data from \textit{Swift}-XRT were incorporated into the broadband SEDs of the sample blazars, using our own analysis of the data. We used the UKSSDC XRT-prods Python API\footnote{\url{https://www.swift.ac.uk/API}} \citep{2009MNRAS.397.1177E, 2007A&A...469..379E}, which retrieves high-level spectral data for the source and background regions along with the instrument response functions. Spectral parameters and flux points were then extracted with \texttt{XSPEC} \citep{1996ASPC..101...17A}. In the optical/UV range, we also included data from the UV/Optical Telescope aboard the Neil Gehrels \textit{Swift} Observatory \citep[\textit{Swift}-UVOT,][]{2005SSRv..120...95R}, which we analysed using an automated tool built on the official HEASARC \texttt{uvot-product} package.

In the HE $\gamma$-ray band, we included flux data from \textit{Fermi}-LAT available in the 4FGL-DR4 catalogue for eight energy bins. The only blazar in our sample detected in the VHE $\gamma$-ray band is 4FGL J0013.9$-$1854 (associated with SHBL J001355.9$-$185406), for which we also included the H.E.S.S. data provided in the Spectral TeV Extragalactic catalogue \citep[STeVECat;][]{2023arXiv230400835G}.  

Additionally, for each source, we incorporated multi-wavelength archival data from the Space Science Data Center (SSDC) SED Builder service\footnote{\url{https://tools.ssdc.asi.it/SED/}}, an online tool for downloading data collected by multiple instruments over several decades. The SSDC SED Builder was used to extract historical data for each blazar across all bands except $\gamma$ rays. Since our study focuses on low-variability sources, we did not apply any time constraints on the SSDC data to exclude potential flaring periods.

\section{Methodology: Broadband SED modelling}
\label{sec:modeling} 

\subsection{Host galaxy emission}

\begin{figure}
	\includegraphics[width=\columnwidth]{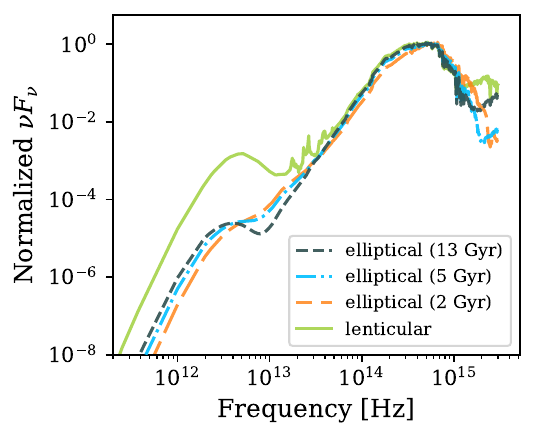}
    \caption{Templates from the SWIRE Template Library used for modelling the host galaxy emission, corresponding to elliptical galaxies of 2 Gyr, 5 Gyr, and 13 Gyr, and to a lenticular galaxy.}
    \label{fig:host_gal}
\end{figure}

Since EHSP blazars have low non-thermal flux at optical/UV energies and their synchrotron peak occurs at higher frequencies, the host galaxy’s thermal emission is often prominent in the optical range of their SEDs. Therefore, before modelling the broadband SEDs, we fitted different host galaxy models to the data in the optical range to determine the best-fitting model. For this purpose, we used four host galaxy templates from the SWIRE Template Library\footnote{\url{http://www.iasf-milano.inaf.it/~polletta/templates/swire_templates.html}} \citep{2007ApJ...663...81P}. The host galaxies of most blazars are elliptical galaxies, characterised by old stellar populations and low star formation rates \citep{2000ApJ...532..816U, 2021ApJS..253...46P}. Lenticular galaxies, while distinct in structure, share these same characteristics and are indistinguishable from ellipticals in the optical band, specifically in the range where we fit the host galaxy templates ($7 \times 10^{13}$ Hz to $10^{15}$ Hz). For this reason, the templates that we employed, shown in Figure~\ref{fig:host_gal}, correspond to elliptical galaxies of 2 Gyr, 5 Gyr, and 13 Gyr, and to a lenticular galaxy. The best-fit host galaxy model was considered to be the one for which the lowest $\chi^2$ value is obtained in the optical range ($7 \times 10^{13}$ Hz to $10^{15}$ Hz). We then assumed black-body emission and, by applying Wien's displacement law, estimated the approximate value of the temperature of the host galaxy. The host galaxy emission is generally not considered a significant source of seed photons for the external Compton process, and therefore, its contribution to the external Compton emission at high energies can be considered negligible. Thus, this contribution was not included in the broadband SED modelling. External Compton emission can also arise from the IR photons of the dusty torus. However, for EHSPs, this contribution is usually negligible, and, therefore, we excluded it from our modelling.

\subsection{Synchrotron and inverse Compton emission}

Once the host galaxy template that best fits the optical data was identified, the broadband SED modelling was conducted using a one-zone synchrotron/SSC model combined with the best-fit host galaxy model and all the data described in Section \ref{sec:data}. In the absence of flaring states, the SEDs of HSPs and extreme blazars are well described by a simple one-zone SSC model \citep[e.g.][]{2022Galax..10..105S}, where the higher energy peak results from inverse Compton scattering of the electron population with photons produced by the synchrotron process. For these blazars, the external Compton component is typically negligible due to the relatively simple environments surrounding their jets, which lack a dusty torus or a BLR to supply the photons necessary for the external Compton mechanism \citep[e.g.][]{2011MNRAS.414.2674G, 2014A&ARv..22...73F}. Therefore, a one-zone SSC model is generally sufficient to describe their SEDs \citep{2008MNRAS.385..283C}.

\subsection{Model assumptions and fitting}

In the one-zone SSC model, both synchrotron and inverse Compton components are assumed to originate from the same region. For the source geometry, we adopted the simplest scenario where the emission is produced in a single spherical region or blob of radius $R$, located within the jet at a distance $R_H$ from the central SMBH. This region is filled with ultra-relativistic electrons moving with a bulk Lorentz factor $\Gamma$. In blazars, the observed emission is highly beamed and Doppler-boosted due to the blob’s relativistic motion and the jet’s small angle relative to our line of sight. The magnetic field, $B$, is considered homogeneous within the blob. 

The electron population was modelled with a smooth broken power law distribution: a lower energy population with spectral slope $p_1$ and Lorentz factors between $\gamma_{\text{min}}$ and $\gamma_{\text{break}}$, and a higher energy population with spectral slope $p_2$ and Lorentz factors between $\gamma_{\text{break}}$ and $\gamma_{\text{max}}$. The smooth broken power law provides an optimal balance between model flexibility and physical interpretability, making it a widely used framework for characterising HSPs and EHSPs \citep[e.g.][]{2015MNRAS.450.4399A, 2017ApJ...836...83S}. To test the robustness of this choice, we applied two alternative electron energy distributions, i.e. a power law with an exponential cut-off and a log-parabola, to a subsample of the sources. Both alternatives yield results consistent with those obtained using the smooth broken power law, indicating that the choice of this electron energy distribution does not introduce significant bias in the modelling. During the fitting, following the methodology from \citet{NIEVAS2022}, $\gamma_{\text{min}}$ was fixed at $10^3$, while $\gamma_{\text{max}}$, $\gamma_{\text{break}}$, $p_1$, and $p_2$ were treated as free parameters. This approach is commonly adopted in modelling studies when low-energy data are ambiguous \citep[e.g.][]{2023ApJS..266...37A, 2025ApJ...980...19O}. To assess the impact of this choice, we repeated the SED modelling on a subsample of sources with $\gamma_{\text{min}}$ left as a free parameter. A comparison of the results obtained with fixed and free $\gamma_{\text{min}}$ reveals no systematic differences, confirming that fixing $\gamma_{\text{min}}$ to $10^3$ does not introduce significant bias into the model.

The SED modelling was performed using the \texttt{JetSeT v1.3.0} package \citep{2020ascl.soft09001T, 2011ApJ...739...66T, 2009A&A...501..879T}, first running the \texttt{lsb} (least square bound) minimiser followed by the \texttt{minuit} minimiser. To reduce the degrees of freedom in the model, some parameters were fixed to typical values for EHSP-like blazars, following \citet{NIEVAS2022}. The radius of the emitting region was set to $R = 10^{16}$ cm, the distance of the blob from the central black hole was set to $R_H = 2 \times 10^{18}$ cm, and the bulk Lorentz factor was fixed at $\Gamma = 20$. Adjusting these parameters impacts SED modelling: a larger emitting region lowers radiation energy density, affecting synchrotron and inverse Compton peaks; changing the SMBH distance alters jet dynamics and energy transfer; varying the Lorentz factor modifies observed flux through Doppler boosting. Despite these sensitivities, the parameters reflect standard EHSP models and general high-frequency blazar behaviour. With these assumptions, seven free parameters remained in the model: the magnetic field strength, $B$; the electron density, $N$; the spectral indices of the low- and high-energy particles, $p_1$ and $p_2$; the maximum Lorentz factor , $\gamma_{\text{max}}$; the break Lorentz factor , $\gamma_{\text{break}}$; and the jet viewing angle , $\theta$. The Doppler boosting factor, $\delta$, was also treated as a free parameter, as it depends on the jet viewing angle, $\theta$. Its dependence on the bulk Lorentz factor and the jet viewing angle is given by

\begin{equation}
    \delta = \frac{1}{\Gamma (1-\beta \cos \theta)} \:, \quad \beta = v/c \:,
\end{equation}
where $\beta$ is the speed of the jet in terms of the speed of light. The parameters used in the SED modelling are listed in Table \ref{table:fit_range}, along with their corresponding fit ranges and whether they were treated as free or frozen during the fitting process.

\begin{table}
\centering
\caption{Parameters used in the SED modelling.}
\label{table:fit_range}
\begin{tabular}{ccc}
\toprule
             Parameter & Free/Frozen &                       Fit Range/Value \\
\midrule
                     R &      frozen &                    $10^{16}$ cm \\
        R$_{\text{H}}$ &      frozen &            $2\times 10^{18}$ cm \\
              $\Gamma$ &      frozen &                              20 \\
                     N &        free &         $[10^{-7},\ 10^7]$ cm$^{-3}$ \\
                     B &        free &               $[10^{-7},\ 100]$ G \\
              $\theta$ &        free &              $[10^{-6},\ 12]$ deg \\
 $\gamma_{\text{min}}$ &      frozen &                        $10^{3}$ \\
 $\gamma_{\text{max}}$ &        free & $[10^4,\ 2\times 10^9]$ \\
$\gamma_{\text{break}}$ &        free &                   $[10,\ 10^9]$ \\
                 $p_1$ &        free &                  $[10^{-6},\ 10]$ \\
                 $p_2$ &        free &                  $[10^{-6},\ 10]$ \\
\bottomrule
\end{tabular}
\tablefoot{The table lists whether each parameter was frozen or free during the fit, and the corresponding range over which fitting was performed.}
\end{table}

\begin{figure}
	\includegraphics[width=\columnwidth]{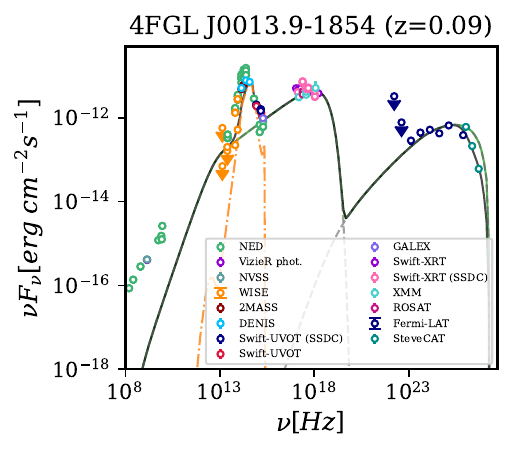}
    \caption{Multi-wavelength emission of blazar 4FGL J0013.9$-$1854 modelled using a one-zone SSC model and a 13 Gyr old elliptical galaxy template for the host. The fluxes used to reconstruct the SEDs are shown as coloured points. The synchrotron and SSC emission from the best-fit model (dashed grey line), the host galaxy emission (orange), the total intrinsic emission summing all components (solid green line), and the observed emission after accounting for EBL absorption (solid black line) are displayed.}
    \label{fig:mwl_sed}
\end{figure}

We fitted the data in the range from $5 \times 10^{10}$ Hz to $10^{27}$ Hz, excluding the radio emission, as it is often believed to originate from a region much larger than the compact region responsible for the rest of the broadband SED. Additionally, at radio frequencies, synchrotron self-absorption becomes significant and may suppress low-frequency emission \citep[e.g.][]{1998ApJ...509..608T}. The absorption of $\gamma$ rays by the EBL was taken into account, with attenuation becoming significant above $\sim$100 GeV for nearby sources but affecting progressively lower energies at higher redshifts, starting around $\sim$30 GeV for $z \sim 1$. The EBL, composed of the accumulated and redshifted light from all galaxies, stars, and interstellar dust in the universe, attenuates $\gamma$ rays through pair-production interactions with high-energy photons. This attenuation was incorporated using the optical depths from \cite{2024MNRAS.527.4632D}, based on the EBL model by \cite{2021MNRAS.507.5144S}. An example of the broadband SED of one source from the sample fitted with this model is shown in Figure~\ref{fig:mwl_sed}. All multi-wavelength SEDs of the 124 sample sources with their best-fit models are presented in Appendix \ref{appendix:mwl_plots}.

Even after applying cuts to the initial sample to exclude potential variable sources in the X-ray band, when considering data from different instruments some sources still show flux dispersion. This variability in X-ray flux could have influenced the SED modelling. To ensure reliable results for the EHSP blazar properties derived here, we excluded sources with poor fitting results by selecting only those with $\chi^2/\text{dof} < 1.5$. The $\chi^2$ was calculated using only the X-ray and $\gamma$-ray bands, considering all data points above $10^{16}$ Hz, since the radio and optical ranges show more dispersion. In total, 113 sources meet this requirement, resulting in the exclusion of 11 sources. These 11 sources were not considered when deriving the general properties of our blazar sample, as shown in Figures~\ref{fig:CD_B},~\ref{fig:CD},~\ref{fig:comparison_plots},~\ref{fig:comparison_CD},~\ref{fig:CD_host_gal},~\ref{fig:energy_budget}, and ~\ref{fig:CD_UBUe}. In addition, to assess the potential bias introduced by limited X-ray coverage, we divided the full sample into two subsamples based on whether the obtained synchrotron peak frequency lay above the highest-energy available X-ray data. The comparison of their key properties revealed no systematic differences (aside from the expected shift in $\nu_{sync}^{peak}$), indicating that the absence of X-ray constraints on the peak position does not introduce a significant bias in our results.

\section{Results and discussion}
\label{sec:results}

\subsection{Broadband SED modelling results}
\label{sec:sed_modeling_results}

The best-fit parameters resulting from the SED modelling performed for each source in the sample are provided in Appendix \ref{appendix:tables}, Table \ref{table:bestfit}\footnote{We make publicly available at Zenodo all the SED fluxes and the results from the SED fits (doi: \href{https://zenodo.org/records/15778400}{10.5281/zenodo.15778399}) and at \url{https://www.ucm.es/blazars/ehsp}. See Appendix \ref{appendix:fits_table} for a description of the structure of the FITS file.}. The SED modelling also allowed us to calculate the synchrotron peak frequency, $\nu_{sync}^{peak}$, and the Compton dominance (CD), which are listed in Table \ref{table:energy-budget} for each source. CD is a parameter used in the study of blazars and other AGNs to describe the relative strength of inverse Compton emission compared to synchrotron emission in their spectral energy distributions. It plays a crucial role in understanding the physical environment of the jet, including the balance of magnetic and particle energy densities, the photon fields driving the inverse Compton process, and the overall energy dissipation mechanisms in the blazar \citep[e.g.][]{2021ApJS..253...46P}.

We find that out of the 113 sources in the sample with good fits, 66 have $\nu_{sync}^{peak} \geq 10^{17}$ Hz and could therefore be classified as EHSPs according to our results. Of the remaining sources, 41 have $10^{15}$ Hz $\leq \nu_{sync}^{peak} < 10^{17}$ Hz and could be classified as HSPs. However, six sources have $10^{14}$ Hz $\leq \nu_{sync}^{peak} < 10^{15}$ Hz, placing them in the ISP class of blazars. These sources are: 4FGL J0146.9$-$5202, 4FGL J1241.8$-$1456, 4FGL J1311.0+0034, 4FGL J1340.8$-$0409, 4FGL J1440.9+0609, and 4FGL J1534.8+3716. The redshift distributions of EHSPs and HSPs/ISPs show no significant differences, providing no clear evidence of an evolutionary connection in our sample \citep{2003A&A...401..927B, 2014ApJ...780...73A}. However, this result may be influenced by redshift uncertainties, particularly those from photometric estimates.

For the jet viewing angle, we obtain values of $\theta < 12^\circ$, with 85\% of the sources having $\theta < 8^\circ$. This corresponds to Doppler factors between $\sim 2$ and 40, typical for blazars due to their jets being closely aligned with our line of sight. For the maximum Lorentz factor $\gamma_{\text{max}}$, we find values ranging from $2.9 \times 10^5$ to $4.6 \times 10^8$. Determining $\gamma_{\text{max}}$ is challenging due to the lack of measurements between $10^{18}$ Hz and $\sim 10^{20}$ Hz, the gap between NuSTAR and \textit{Fermi}-LAT energy ranges, which may result in $\gamma_{\text{max}}$ values larger than expected.

\begin{figure}
	\includegraphics[width=\columnwidth]{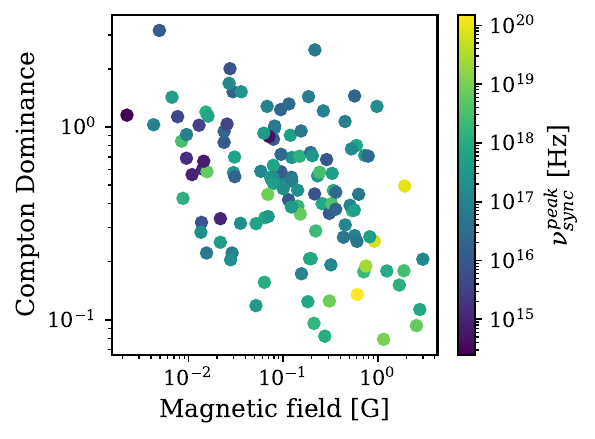}
    \caption{Compton dominance as a function of the magnetic field for the 113 sources in the sample with a good fitting result. The colour scale represents the synchrotron peak frequency.}
    \label{fig:CD_B}
\end{figure}

\begin{figure}
    \centering
	\includegraphics[width=\columnwidth]{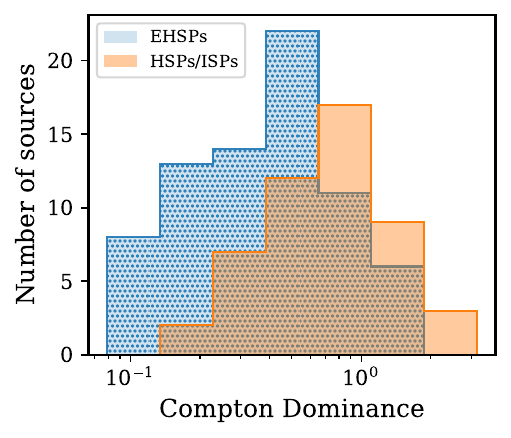}
    \caption{Compton dominance distribution of the sources in the sample with $\nu_{sync}^{peak}\geq 10^{17}$ Hz (EHSPs, in blue) and $\nu_{sync}^{peak} < 10^{17}$ Hz (HSPs/ISPs, in orange).} 
    \label{fig:CD}
\end{figure}

Figure~\ref{fig:CD_B} shows the relationship between CD, magnetic field strength, and synchrotron peak frequency for the 113 sources with good fitting results. Sources with lower synchrotron peak frequencies generally have lower magnetic field values and higher CD, while the most extreme sources (with the highest synchrotron peak frequencies) display higher magnetic field values and lower CD. Figure \ref{fig:CD} presents the CD distribution for the EHSP and HSP/ISP candidates in our sample, showing that EHSP candidates typically have lower CD values than less energetic sources, as demonstrated by \citet{2021ApJS..253...46P}. However, all sources in our sample have modest CD, with the highest being CD $=3.2$, corresponding to an HSP source (4FGL J1419.3$+$0444, $\nu_{sync}^{peak} = 1.0 \times 10^{16}$ Hz) based on our results. In contrast, the EHSP candidate with the highest CD value has CD $=1.5$ (4FGL J1719.3$+$1205).

\begin{figure*}
	\includegraphics[width=\textwidth]{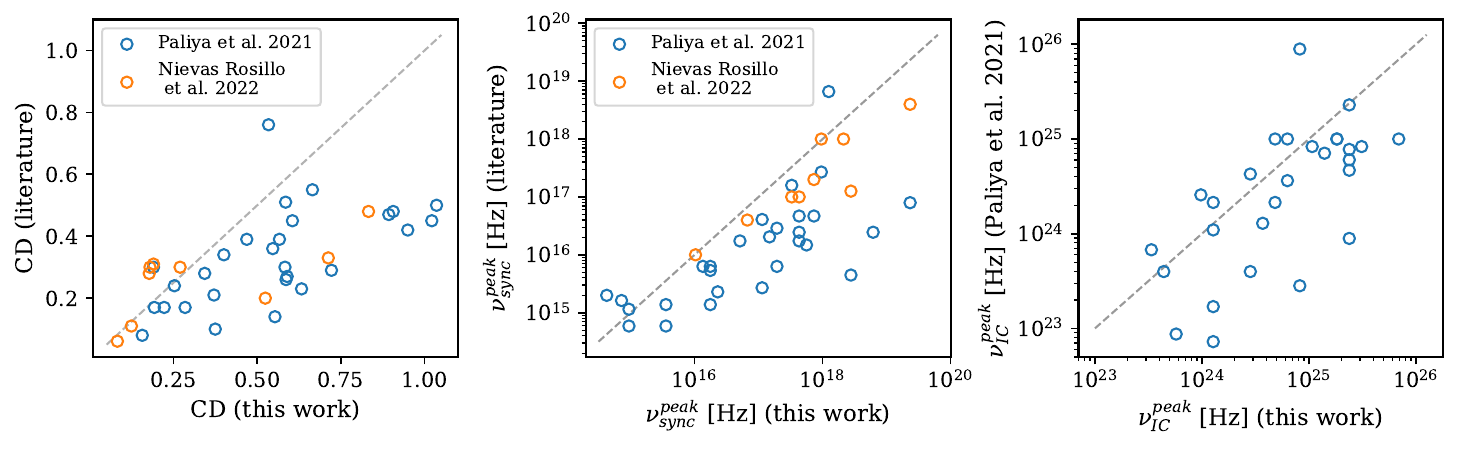}
    \caption{Comparison of the CD, synchrotron peak frequency, and inverse Compton peak frequency obtained in this work with those reported by \citet{2021ApJS..253...46P} for the 28 overlapping sources. The synchrotron peak frequency $\nu_{sync}^{peak}$ and CD obtained in this work are also compared with the $\nu_{sync}^{peak}$ and CD values from \citet{NIEVAS2022} for the nine sources present in both samples. The dashed grey line indicates the one-to-one correspondence.}
    \label{fig:comparison_plots}
\end{figure*}

Out of the 113 sources with good fits, 28 overlap with those in \citet{2021ApJS..253...46P}, which catalogues 1077 $\gamma$-ray blazars or BCUs detected by \textit{Fermi}-LAT with available optical spectra or measurements of host galaxy bulge magnitude, central SMBH mass, or disc luminosity. Figure~\ref{fig:comparison_plots} compares our results for CD and the synchrotron ($\nu_{sync}^{peak}$) and inverse Compton ($\nu_{IC}^{peak}$) peak frequencies for these 28 sources with those from \citet{2021ApJS..253...46P}. Their model does not account for host galaxy emission, potentially increasing the modelled non-thermal synchrotron component. This could partly explain why our $\nu_{sync}^{peak}$ values are systematically higher by a factor around 6, although a small contribution may arise from the inherent uncertainty introduced by the manual fitting procedure and differences between the criteria adopted in each work, the data selection or the modelling codes.

This difference in $\nu_{sync}^{peak}$ directly impacts CD, which depends on the balance between synchrotron and inverse Compton emissions. Our CD values are consistently higher than those in \citet{2021ApJS..253...46P}, reflecting the weaker synchrotron component. In contrast, $\nu_{IC}^{peak}$ shows less deviation between the studies, suggesting it is less sensitive to modelling assumptions. However, this stability in $\nu_{IC}^{peak}$ does not offset the influence of the higher $\nu_{sync}^{peak}$ on CD.

Additionally, Figure~\ref{fig:comparison_plots} compares our $\nu_{sync}^{peak}$ and CD values with those from \citet{NIEVAS2022}, based on 22 2BIGB catalogue sources classified as BCU in 4FGL. Nine sources overlap between the two samples, showing better agreement than with \citet{2021ApJS..253...46P}, likely due to \citet{NIEVAS2022} including a blackbody component to account for host galaxy emission. Still, our $\nu_{sync}^{peak}$ values remain systematically a factor of 3 higher, highlighting the existing systematic uncertainties arising from the modelling techniques used in these works.

\begin{figure}
	\includegraphics[width=\columnwidth]{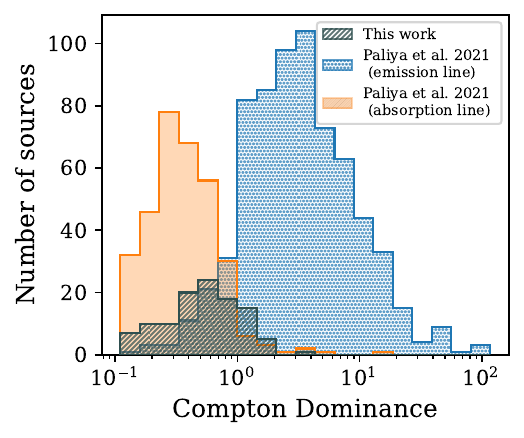}
    \caption{Comparison of the CD distribution extracted from our results for the 113 sources in the sample with good fits to the results from \citet{2021ApJS..253...46P} for emission-line and absorption-line blazars.}
    \label{fig:comparison_CD}
\end{figure}

In Figure \ref{fig:comparison_CD}, the CD distribution of our sample sources is shown, compared with that of the emission-line and absorption-line blazars from \citet{2021ApJS..253...46P}. The sources selected for this work appear to lie at the low end of the absorption-line blazar sample, suggesting that our sources can be classified within this subset of blazars. This population primarily consists of BL Lacs, whose spectra show absorption lines attributed to the stellar population of the host galaxy, while the emission-line blazars are typically identified as FSRQs. Note that blazar classification as absorption-line or emission-line objects depends on the source state, as even BL Lacs can display emission lines when the jet is not in a high state. Moreover, our predominantly EHSP sample shows higher CD values than their absorption-line blazars, suggesting greater radiative efficiency and stronger intrinsic absorption. While such absorption would typically shift the synchrotron peak to lower frequencies, our sources maintain high synchrotron peak values.

Furthermore, a correlation between accretion luminosity ($L_{\text{disc}}$) and CD in blazars was found by \citet[][shown in Figure 11 of that work]{2021ApJS..253...46P}, suggesting that blazars with high CD tend to be more luminous. In particular, emission-line blazars generally have CD $> 1$ and $L_{\text{disc}}/L_{\text{Edd}} > 0.01$, while absorption-line blazars typically have CD $< 1$ and $L_{\text{disc}}/L_{\text{Edd}} < 0.01$. Based on these findings and given that most of the EHSP candidates in our sample show CD values of $\text{CD} \lesssim 1$, we expect these blazars to generally have $L_{\text{disc}}$ values in Eddington units of $L_{\text{disc}} / L_{\text{Edd}} \leq 0.01$. Therefore, they can be classified as low-Compton-dominated (LCD) objects, according to the classification proposed by \citet{2021ApJS..253...46P}. In contrast, more Compton-dominated blazars, primarily FSRQs, are typically classified as high-Compton-dominated (HCD) objects. Hence, we conclude that our blazar sample mainly consists of LCD objects, with no emission lines and low accretion activity.

The FSRQ and the radiogalaxy in our sample, 4FGL J0132.7$-$0804 (PKS 0130$-$083) and 4FGL J1518.6+0614 (TXS 1516+064), respectively, both have CD $< 1$. Specifically, the results obtained are CD $= 0.27$ for 4FGL J0132.7$-$0804 and CD $= 0.12$ for 4FGL J1518.6+0614. Following the correlation derived in \citet{2021ApJS..253...46P}, these low CD values suggest that these two sources have low accretion activity or that the emission site is located far from the core (i.e. the strong radiation fields have already weakened by the time they reach the distant emitting region), although FSRQ blazars typically have CD $\gtrsim 1$. However, in the optical spectrum of 4FGL J0132.7$-$0804 shown in Figure 4 of \citet{2021AJ....162..177P}, the emission lines H$\alpha$ and H$\beta$ appear to be very narrow, suggesting that this source is not a broad-line blazar or FSRQ, but rather a BL Lac. Therefore, the low CD of this source may be due to a weaker beaming effect rather than low accretion activity.

Out of the 113 sample sources with good fitting results, 28 are classified as BCUs in the 4FGL catalogue. According to our modelling results, all but two of these BCUs have CD $<1$, showing similar emission properties to those of BL Lacs. The remaining two BCUs, for which we obtain CD $\geq 1$, are 4FGL J0611.1+4325 (with CD $=1.4$) and 4FGL J1719.3+1205 (with CD $=1.5$).

An analysis of potential correlations between black hole mass and key jet properties, including jet viewing angle, magnetic field strength, energy density ratios, jet power components, synchrotron peak frequency, and CD, for the sources shared with \citet{2021ApJS..253...46P} revealed no significant trends. This suggests that jet emission in EHSPs is primarily governed by local jet conditions rather than the mass of the central black hole. Moreover, EHSPs consistently show extreme synchrotron peaks and low CD across the full range of black hole masses. 

\subsection{Host galaxy classification}

After modelling all the sources in the sample, we find that the best-fit host galaxy model for 59 sources corresponds to the lenticular galaxy template, while for 54 sources, it corresponds to an elliptical galaxy. Among these, 27 sources match an elliptical galaxy of 13 Gyr, ten sources match an elliptical galaxy of 5 Gyr, and 17 sources match an elliptical galaxy of 2 Gyr. We note that, although the fitting was done using four different host galaxy templates, it is often difficult to distinguish between the three elliptical galaxy templates, as seen in Figure \ref{fig:host_gal}. The best-fit host galaxy results given in Table \ref{table:bestfit} correspond to the host galaxy template that yielded the lowest $\chi^2$ value, but the difference between the different elliptical galaxy templates can be minimal. For the temperature of the host galaxy, we obtain values ranging from $3.2 \times 10^3$ K to $6.4 \times 10^3$ K.

\begin{figure}
    \centering
	\includegraphics[width=\columnwidth]{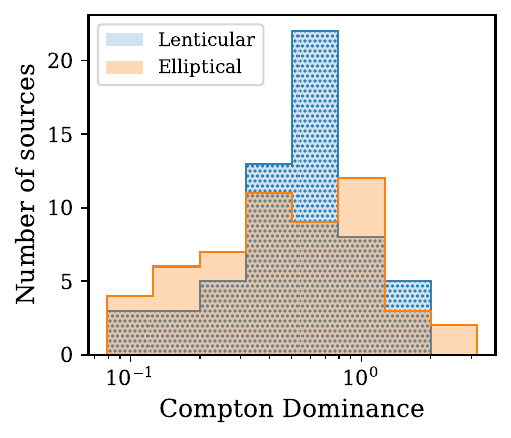}
    \caption{Compton dominance distribution of the resulting lenticular (blue) and elliptical host galaxies (orange) in the sample.}
    \label{fig:CD_host_gal}
\end{figure}

In our broadband SED modelling, we used templates for elliptical and lenticular galaxies to estimate the thermal emission from the host galaxy. Figure~\ref{fig:CD_host_gal} shows that no significant differences are observed between the two types of galaxies. This supports the conclusion that the emission from the host galaxy has a negligible impact on the blazar's non-thermal emission. Even in galaxies with high star formation rates, the thermal emission is unlikely to significantly contribute to the high-energy regime, particularly for EHSPs. This is consistent with the findings of \citet{2000ApJ...532..740S} and \citet{2011MNRAS.414.2674G}, who argued that the thermal emission from the host galaxy is generally not a significant source of seed photons for the external Compton mechanism. While external Compton on star-forming photons could contribute additional components in some extreme cases, like M82 \citep{2017MNRAS.469..255G}, we find that this effect does not substantially alter the SEDs of the majority of EHSPs.

\subsection{Energy budget}

\begin{figure*}
	\includegraphics[width=\textwidth]{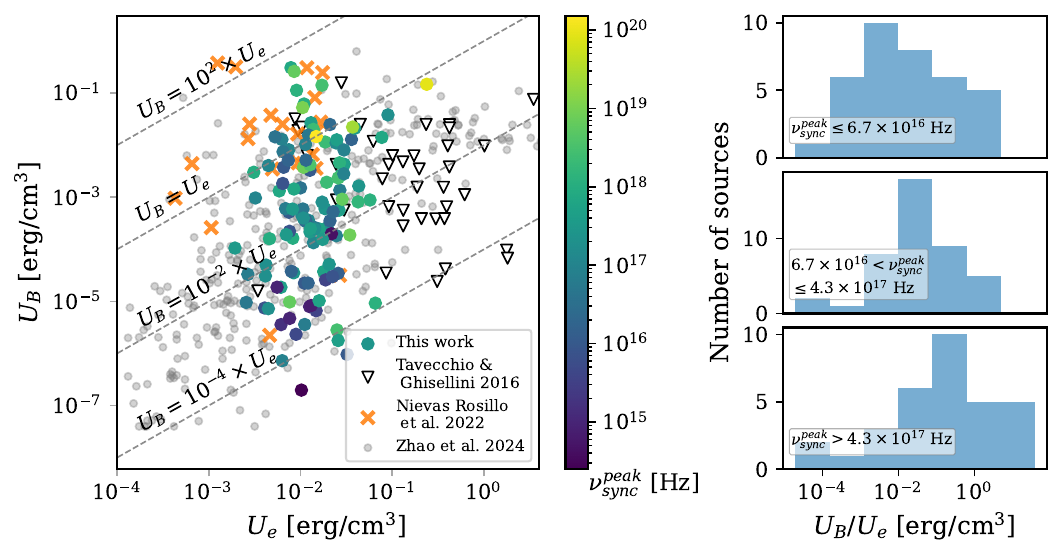}
    \caption{Left: Magnetic energy density versus kinetic energy density for the 113 sources in the sample with a good fitting result, with colour representing their synchrotron peak frequency. For comparison, the results obtained by \citet{TAVECCHIO2016} (as black triangles), \citet{NIEVAS2022} (as orange crosses), and \citet{ZHAO2024} (as grey circles) are shown. The dashed grey lines represent the lines for $U_B = 10^2 \times U_e$, $U_B = U_e$ (equipartition), $U_B = 10^{-2} \times U_e$, and $U_B = 10^{-4} \times U_e$. Right: Distribution of the ratio of magnetic energy density to kinetic energy density ($U_B/U_e$) for three different samples, corresponding to three ranges of synchrotron peak frequency: $\nu_{\text{sync}}^{\text{peak}} \leq 6.7 \times 10^{16}$ Hz (top), $6.7 \times 10^{16}$ Hz $< \nu_{\text{sync}}^{\text{peak}} \leq 4.3 \times 10^{17}$ Hz (middle), and $\nu_{\text{sync}}^{\text{peak}} > 4.3 \times 10^{17}$ Hz (bottom).}
    \label{fig:energy_budget}
\end{figure*}

\texttt{JetSeT} modelling also provides an energy report for the resulting best-fit model, including the magnetic and kinetic energy densities, $U_B$ and $U_e$, respectively. The former quantifies the energy stored in the magnetic field per unit volume, while the latter represents the energy carried by the population of relativistic electrons, often derived from the modelled electron distribution in the jet. Additional outputs from \texttt{JetSeT} include the luminosity of each emission process, i.e. synchrotron and IC, and the jet luminosity associated with radiative mechanisms, electrons, and the magnetic field. The ratio $U_B/U_e$, representing the energy density of the magnetic field to that of the relativistic electrons, is a critical parameter for understanding the physical conditions and energy balance within blazar jets. This quantity reveals the jet's magnetisation and the balance between magnetic fields and particle energies in producing the observed radiation. A jet close to equipartition ($U_B / U_e \sim 1$) is thought to be energetically efficient, as such a configuration minimises energy losses during the acceleration and transport of particles. In the context of EHSPs, studying $U_B / U_e$ helps to probe the mechanisms responsible for their extreme synchrotron peak frequencies, such as efficient particle acceleration and strong magnetic fields. Furthermore, deviations from equipartition can indicate variations in jet dynamics, such as regions of high particle dominance during flaring states or magnetic dominance in steady emission. These characteristics are crucial for modelling the jets and understanding the interplay between synchrotron and inverse Compton processes, as well as the overall energy budget of the system. These energy budget results are given in Table \ref{table:energy-budget}. Figure \ref{fig:energy_budget} illustrates the relationship between the magnetic and kinetic energy densities resulting from the SED modelling for the 113 sources in the sample with a good fitting result ($\chi^2/\text{dof} < 1.5$). For comparison, the results for the magnetic and kinetic energy densities from \citet{TAVECCHIO2016}, \citet{NIEVAS2022}, and \citet{ZHAO2024} are also shown.

Interestingly, our results suggest a relation between the $U_B/U_e$ ratio and the synchrotron peak frequency, with the least energetic sources having lower values of $U_B/U_e$, and the most extreme sources located closer to the line $U_B \approx U_e$. To verify if there is a dependence of the $U_B/U_e$ ratio on the synchrotron peak frequency, we divided our sample into three subsamples with different ranges of $\nu_{sync}^{peak}$ values obtained from our SED modelling. The three subsamples have approximately the same number of sources, and their $U_B/U_e$ distributions are shown in Figure \ref{fig:energy_budget}. We fitted each of the three distributions of the $U_B/U_e$ ratio to a Gaussian function, obtaining the following results for the mean and standard deviation: for the distribution of sources with $\nu_{sync}^{peak} \leq 6.7 \times 10^{16}$ Hz, $\langle \log (U_B/U_e) \rangle = -1.89$, $\sigma = 1.30$; for the distribution of sources with $6.7 \times 10^{16}$ Hz $< \nu_{sync}^{peak} \leq 4.3 \times 10^{17}$ Hz, $\langle \log (U_B/U_e) \rangle = -1.45$, $\sigma = 1.08$; and for the distribution of sources with $\nu_{sync}^{peak} > 4.3 \times 10^{17}$ Hz, $\langle \log (U_B/U_e) \rangle = -0.90$, $\sigma = 1.40$. Therefore, we conclude that the most extreme sources are generally closer to equipartition than the sources with lower synchrotron peak frequencies. This is in agreement with the results obtained by \citet{NIEVAS2022}, where most of the sources in their sample, which contains 17 EHSP candidates out of the total 22 sources, were close to equipartition.

However, most sources from the samples of \citet{ZHAO2024} and \citet{TAVECCHIO2016} are far from equipartition, with the magnetic energy density being much smaller than the electron energy density, clustering mainly around the line $U_B = 10^{-2} \times U_e$. The source sample used by \citet{ZHAO2024} contains 348 HSP blazars. Their initial sample was selected by collecting all the blazars classified as HBL in the 4FGL catalogue. Afterwards, they modelled the broadband SED and selected only those blazars with a synchrotron peak frequency of $\nu_{\text{sync}}^{\text{peak}} \geq 10^{15}$ Hz. Among these resulting 348 sources, 42 are also included in our sample. On the other hand, \citet{TAVECCHIO2016} selected 45 BL Lac objects, 12 of which were detected in the tera-electronvolt $\gamma$-ray band. Note that sources detected in the tera-electronvolt band may show biased results since they are typically detected during flaring episodes. These sample selections can help explain the differences in the obtained results, since the source samples from \citet{TAVECCHIO2016} and \citet{ZHAO2024} also include variable sources. This variability suggests that during certain observations, such sources may be far from equilibrium (e.g. during flares), and in those cases, the electron energy injection could be higher. This would lead to a lower magnetisation, and consequently, to a lower $U_B/U_e$ ratio. In contrast, the sources in our sample are characterised by low variability across various wavelengths, due to the selection criteria applied, and it is expected that they are closer to equipartition.

\begin{figure}
	\includegraphics[width=\columnwidth]{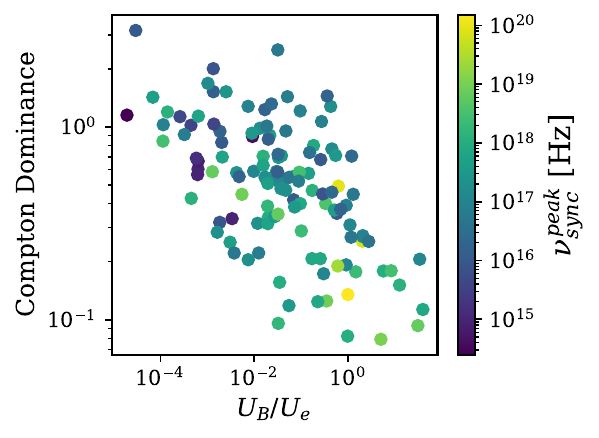}
    \caption{Compton dominance as a function of the ratio between the magnetic and the kinetic energy density.}
    \label{fig:CD_UBUe}
\end{figure}

Figure \ref{fig:CD_UBUe} illustrates the correlation between the CD and the magnetic to kinetic energy density ratio, showing that sources closer to equipartition tend to have lower CD values. This trend supports the idea that EHSPs typically lack strong external photon fields, and are dominated by SSC emission rather than external Compton processes. The low Compton dominance in these sources indicates that inverse Compton scattering is relatively inefficient, likely due to both weak external fields and a near-equipartition state within the jet. Notably, the observed trend also suggests that more magnetically dominated jets tend to produce less Compton-dominated emission, highlighting the connection between jet magnetisation and the radiative output of the jet.

To test whether the trends observed with synchrotron peak frequency persist when using a more intrinsic physical parameter, we computed for each source the characteristic Lorentz factor, $\gamma_{3P}$, defined as the value that maximises $\gamma^3 n(\gamma)$ in the best-fit electron energy distribution. While some correlations, such as with the magnetic field or Compton dominance, are also observed with $\gamma_{3P}$, others, most notably the trend between $U_B/U_e$ and $\nu_{sync}^{peak}$, are not evident when using $\gamma_{3P}$. This suggests that $\nu_{sync}^{peak}$ reflects a more complex combination of parameters (including $\gamma$, B, and $\delta$) and may capture broader physical conditions in the jet.

\subsection{The blazar sequence and the role of EHSP blazars}

The blazar sequence, a widely used framework in blazar astrophysics, posits an anti-correlation between luminosity and synchrotron peak frequency, with FSRQs typically representing luminous, low synchrotron peak sources, and HSPs and EHSPs occupying the lower-luminosity, high-peak regime \citep{1998MNRAS.299..433F, 2017MNRAS.469..255G}. Our identification of 66 new EHSP candidates reinforces aspects of this model while raising intriguing questions about its universality. The prevalence of EHSPs at the lower-luminosity end aligns with the expected anti-correlation, while their low CD values, as seen in Figure \ref{fig:CD}, reflect the reduced influence of external photon fields as synchrotron peak frequency increases \citep{2017MNRAS.469..255G}.

However, the large number of EHSPs we identify may challenge previous assumptions about this source type. Instrumental sensitivity limits, biased selection criteria, incomplete data coverage, and theoretical models may not have fully captured the diversity of blazars \citep[e.g.][]{2016Galax...4...36G, 2020NatAs...4..124B}. If EHSPs are more common than previously thought, the sharp dichotomy between low-luminosity, high-peak blazars and their luminous, low-peak counterparts could soften, suggesting a more continuous distribution. This shift challenges the sequence’s traditional boundaries and invites consideration of alternative interpretations. Furthermore, the difficulty of detecting EHSPs at high redshifts due to their lower luminosity introduces a selection bias \citep[e.g.][]{2012MNRAS.420.2899G, 2015ApJ...804...74P, 2022ApJS..263...24A}, potentially skewing the observed relationship between luminosity and synchrotron peak frequency.

The trends in CD provide additional context. The prevalence of low CD values among EHSPs suggests that synchrotron/SSC processes dominate their emission, consistent with the absence of strong external photon fields \citep{2021ApJS..253...46P}. This contrasts with the high CD values typically seen in FSRQs, where external Compton scattering plays a significant role \citep[e.g.][]{2022Galax..10..105S}. Comparing the CD distribution of EHSPs to other blazar types (see Figure \ref{fig:comparison_CD}) highlights the distinct physical environments and emission mechanisms across the blazar population.

With their extreme synchrotron peaks and low CD, EHSPs may represent either the natural extension of the blazar sequence or a distinct subclass that deviates from its predictions. Some of the most extreme EHSPs in the sample, particularly those with the highest peak frequencies and CD $\gtrsim 1$, challenge the traditional framework \citep[e.g.][]{2010MNRAS.401.1570T}. If their underlying jet physics or environmental conditions differ significantly, they could necessitate a revised model that better captures the diversity of the blazar population. The detectability predictions for EHSPs with the CTAO provide an opportunity to test these ideas. Observations of EHSPs at higher redshifts or with unexpected luminosity characteristics could further refine the blazar sequence or point towards its limitations.

These results suggest that while the blazar sequence provides a useful framework, it may oversimplify the diversity of blazars. Blazars of the EHSP class challenge the universality of the sequence, pointing to the need for models that accommodate the complexities of their environments, emission mechanisms, and population statistics \citep[e.g.][]{2020NatAs...4..124B, 2020ApJS..247...16A}. This perspective enriches the understanding of blazars and provides a foundation for exploring their classification and evolution in greater depth.

\section{Detectability predictions with CTAO}
\label{sec:cta-significance}
\begin{table}
\centering
\caption{Expected detection significance with CTAO for sources with $\geq 3\sigma$ in 20 hours.}
\label{table:cta-significance10}
\begin{tabular}{ccccc}
\toprule
     4FGL Name & \thead{CTAO \\ ($\sigma $)} & Redshift & \thead{Redshift \\ reference} & $t_{5\sigma}$ [h] \\
\midrule
J0013.9$-$1854 &                         3.9 &    0.095 &                           P21 &                         34 \\
J0039.1$-$2219 &                         9.8 &    0.062 &                          4LAC &                           -- \\
J0054.7$-$2455 &                         4.9 &     0.12 &                          4LAC &                         21 \\
J0059.3$-$0152 &                         4.3 &     0.14 &                           P21 &                         27 \\
J0123.7$-$2311 &                         5.7 &      0.4 &                           G21 &                           -- \\
J0336.5$-$0348 &                         5.0 &     0.16 &                          4LAC &                           -- \\
J0357.2$-$0319 &                         3.4 &      0.3 &                          3HSP &                         44 \\
J0536.4$-$3343 &                         8.3 &     0.34 &                          4LAC &                           -- \\
J0558.0$-$3837 &                         4.5 &      0.3 &                           P21 &                         24 \\
J0813.7$-$0356 &                         3.3 &     0.33 &                          3HSP &                         47 \\
J0912.9$-$2102 &                        17.0 &      0.2 &                           P21 &                           -- \\
J1132.2$-$4736 &                         3.5 &     0.23 &                           P21 &                         40 \\
J1256.2$-$1146 &                         5.5 &    0.058 &                           P21 &                           -- \\
J1310.2$-$1158 &                         3.3 &     0.14 &                           G21 &                         46 \\
J1539.7$-$1127 &                         4.9 &     0.22 &                          3HSP &                         21 \\
J1656.9$-$2010 &                         5.1 &     0.46 &                           G21 &                           -- \\
J1841.3$+$2909 &                         3.6 &     0.29 &                           G21 &                         39 \\
J1944.4$-$4523 &                         3.9 &     0.21 &                          3HSP &                         34 \\
J2041.9$-$3735 &                         6.1 &    0.099 &                           G21 &                           -- \\
J2340.8$+$8015 &                         6.1 &     0.27 &                           G21 &                           -- \\
\bottomrule
\end{tabular}
\tablefoot{The table gives the expected detection significance with CTAO and the redshift of the 20 sources of the sample that have an expected detection significance $\geq 3 \sigma $ with CTAO in 20 hours of observation. For the sources with an expected detection significance $\geq 3 \sigma $, but $< 5 \sigma$, the time needed to obtain a $5 \sigma$ detection is also given in the fifth column ($t_{5\sigma}$). The sources are sorted by right ascension (RA) in increasing order. In the redshift reference column, P21 corresponds to \citet{2021ApJS..253...46P} and G21 to \citet{goldoni_2021_5512660}.}
\end{table}

Characterised by spectra extending to extremely high frequencies, EHSPs are generally regarded as promising VHE emitters \citep[e.g.][]{2019MNRAS.486.1741F, 2020ApJS..247...16A, 2021ApJ...916...93Z}. Since the higher-energy component of the SED peaks in the VHE range for many EHSPs, studying these sources at tera-electronvolt energies ($\sim 10^{26}$ Hz) is essential to understand the acceleration processes occurring in their jets. However, their low fluxes, relatively steady emission, and EBL attenuation at tera-electronvolt energies make their detection in the VHE band challenging. As a result, only a few EHSPs have been observed above 100 GeV, mostly by the current generation of IACTs: MAGIC, VERITAS, and H.E.S.S. The upcoming Cherenkov Telescope Array Observatory \citep[CTAO;][]{2019scta.book.....C} will represent the next generation of IACTs. These telescopes detect the faint flashes of Cherenkov light produced by the interaction of the $\gamma$ rays with the atmosphere and can only operate during the night, resulting in low duty cycles. Consequently, selecting promising targets for VHE detection is important to optimise the observation time of IACTs. Observations by arrays such as LHAASO or SWGO could also complement these observations for the highest-energy $\gamma$-ray emitters.

Using the spectral shape resulting from the SED modelling explained in Section \ref{sec:modeling}, we estimated the expected detection significance of the modelled sources with the future CTAO assuming a total observation time of 20 hours, following the procedure presented in \citet{2024MNRAS.527.4763D}. For this, our best-fit broadband model was extrapolated to tera-electronvolt energies, incorporating $\gamma$-ray absorption due to EBL interactions using the optical depths from \citet{2024MNRAS.527.4632D}, based on the EBL model by \citet{2021MNRAS.507.5144S}. The significance estimation for each source was done using the CTAO instrument response functions\footnote{\url{https://zenodo.org/records/5499839}} (IRFs) for the `Alpha Configuration' with an average azimuth angle. These IRFs, produced for three different zenith angles (20 degrees, 40 degrees, and 60 degrees), are available for both the north and south hemisphere CTAO arrays. The CTAO-North IRFs were used for sources with positive declination, while the CTAO-South IRFs were used for sources with negative declination. The zenith angle at which each source is observed was determined by assuming observations around culmination time, and the IRF configuration corresponding to the closest zenith angle was used for the significance estimation. Note that the energy threshold is taken as the energy of the lowest bin of the differential sensitivity curve, meaning that the threshold is higher for larger zenith angle observations and for sources in the southern hemisphere. This is because the CTAO northern array, equipped with larger telescopes, is optimised for greater sensitivity to lower-energy $\gamma$-rays, whereas the southern array is designed to detect the highest-energy Cherenkov showers.  
Following the procedure by \cite{2024MNRAS.527.4763D}, point-like differential sensitivity curves for a total exposure time of 20 hours were generated by scaling the IRFs corresponding to a 5-hour exposure with \texttt{gammapy} \citep{2017ICRC...35..766D, acero_fabio_2022_7311399}. From this analysis, we derived the differential flux, as well as the number of excess and background events required to generate a $5 \sigma$ signal within each energy bin. This number of excess events was then scaled linearly with the ratio of the differential fluxes, which is the differential flux in each energy bin for the assumed spectral shape of the source divided by the differential flux derived from the sensitivity curve. As a result, the number of excess events for the considered source was obtained. Finally, the expected detection significance for each source was estimated by applying the equation from \cite{1983ApJ...272..317L} to the obtained number of excess and background events, using the \texttt{WStatCountsStatistic} function in \texttt{gammapy}.

Out of the 113 blazars in the sample, nine sources appear to be potential VHE emitters, with an expected detection significance $\geq 5 \sigma$ after 20 hours of exposure with CTAO. There are 11 additional sources with an expected detection significance $\geq 3 \sigma$ that could be detected with a longer exposure (see Table \ref{table:cta-significance10}). According to our SED modelling results, of the 20 sources detectable above $3 \sigma$, 12 sources have $\nu_{sync}^{peak} \geq 10^{17}$ Hz. Additionally, all these sources are characterised by very low magnetisation, with magnetic field values ranging from 0.0084 to 0.74 Gauss. However, we observe no clear relation between the $U_B/U_e$ ratio and their detectability predictions with the future CTAO, adding to the findings of \citet{NIEVAS2022}, who identified the two potential VHE emitters in their sample as having the smallest $U_B/U_e$ ratios.

Notably, the only source in our blazar sample that has already been detected by an IACT (4FGL J0013.9$-$1854, or SHBL J001355.9$-$185406) shows an expected detection significance of 3.9$\sigma$ in our analysis, calculated for 20 hours of CTAO observations. This result aligns well with its previous detection by H.E.S.S., a less sensitive instrument, which required 41.5 hours of exposure. The source's steady emission across all wavelengths \citep{2013A&A...554A..72H} and the possibility of it undergoing a mild $\gamma$-ray flare during the H.E.S.S. observations further highlight its potential as a strong candidate for VHE detection with CTAO.

\section{Summary and conclusions}
\label{sec:summary}

In this work, we conducted a systematic search for EHSPs by reconstructing and modelling the broadband SEDs of 124 blazars selected from the 2BIGB catalogue, further analysing 113 of them with good fits. To build the multi-wavelength SEDs, an analysis of \textit{Swift}-XRT and \textit{Swift}-UVOT data was performed. In addition, the {\it Fermi}-LAT flux data provided in the 4FGL catalogue were added in the $\gamma$-ray band, as well as archival data from the SSDC SED builder service. Integrating multi-wavelength datasets from different epochs and analyses, involving different settings and methodologies, highlights the importance of open science efforts, and the development of standardised multi-wavelength data formats and analyses.

The broadband SEDs were modelled assuming non-flaring states, allowing us to determine key parameters such as synchrotron peak frequency ($\nu_{sync}^{peak}$) and CD. Using a one-zone SSC model, complemented by host galaxy templates, we identified 66 new EHSP candidates, approximately doubling the known population of these rare and intriguing objects. The results revealed that most EHSPs in the sample present low CD values (CD $< 1$), consistent with SSC-dominated emission in environments with weak external photon fields. Host galaxy emission was fitted using templates for lenticular and elliptical galaxies, but no significant differences were observed in the radiative properties of sources hosted by different galaxy types. This suggests that the thermal contribution from the host galaxy has a negligible impact on the high-energy emission.

The analysis of the energy densities in the jet, $U_B$ (magnetic) and $U_e$ (electron), revealed that EHSPs are generally closer to equipartition ($U_B / U_e \sim 1$) compared to less extreme blazars. This indicates that their jets are radiatively efficient, possibly due to finely balanced particle acceleration and magnetic field strengths. A correlation between $\nu_{sync}^{peak}$ and $U_B / U_e$ suggests that higher synchrotron peak frequencies are associated with jets closer to energy equilibrium. Our results complement and extend the findings of previous works, such as \citet{NIEVAS2022} and \citet{ZHAO2024}, while highlighting differences in the relationship between $U_B / U_e$ and detectability. This underscores the importance of sample selection and variability criteria in shaping the inferred physical properties of EHSPs.

Using the modelled SEDs, we calculated detectability predictions for CTAO. Nine sources were identified as strong candidates for VHE $\gamma$-ray detection, with expected detection significances above $5 \sigma$ in 20 hours of observation. An additional 11 sources could be detected at $5 \sigma$ with slightly longer exposures.

In conclusion, this study demonstrates the value of broadband SED modelling in identifying and characterising EHSPs, providing a better understanding of the extreme physics of their jets. The identification of 66 new candidates not only broadens the population of known EHSPs but also provides critical targets for future observations with CTAO. These sources serve as natural laboratories for exploring particle acceleration, jet dynamics, and the broader blazar sequence. Future multi-wavelength campaigns and deeper VHE observations will be essential to further refine our understanding of these extraordinary objects and their role within the AGN population.

\section*{Data availability}
\label{sec:data_availability}

Tables \ref{table:cta-significance10}, \ref{table:bestfit} and \ref{table:energy-budget} are available at the CDS via \href{http://cdsweb.u-strasbg.fr/cgi-bin/qcat?J/A+A/}{http://cdsweb.u-strasbg.fr/cgi-bin/qcat?J/A+A/}.

\vspace{1em}

\begin{acknowledgements}

        M.L. and J.L.C acknowledge the support of MCIN project PID2022-138172NB-C42 and grant PRE2020-093502.
        
        M.N.R. acknowledges support from the Agencia Estatal de Investigación del Ministerio de Ciencia, Innovación y Universidades (MCIU/AEI) under grant PARTICIPACIÓN DEL IAC EN EL EXPERIMENTO AMS and the European Regional Development Fund (ERDF) with reference PID2022- 137810NB-C22/DIO 10.13039/501100011033.

        A.D. is thankful for the support of Proyecto PID2021-126536OA-I00 funded by MCIN / AEI /10.13039/501100011033.
        
        This research has made use of the CTA instrument response functions provided by the CTA Consortium and Observatory, see \url{https://www.ctao-observatory.org/science/cta-performance/} \citep[version prod5 v0.1;][]{cherenkov_telescope_array_observatory_2021_5499840} for more details. This work made use of data supplied by the UK Swift Science Data Centre at the University of Leicester. Part of this work is based on archival data, software or online services provided by the Space Science Data Center - ASI.

\end{acknowledgements}

\bibliographystyle{aa} 
\bibliography{references} 

\onecolumn

\begin{appendix}

\section{Broadband SEDs and best-fit model of the sample sources}
\label{appendix:mwl_plots}
This section presents the broadband SEDs used to model the 124 sources in our sample, along with the best-fit SSC model. 

\begin{figure}[H] 
\centering 

    \begin{subfigure}[b]{\textwidth} 
        \centering 
        \includegraphics[width=0.9\textwidth]{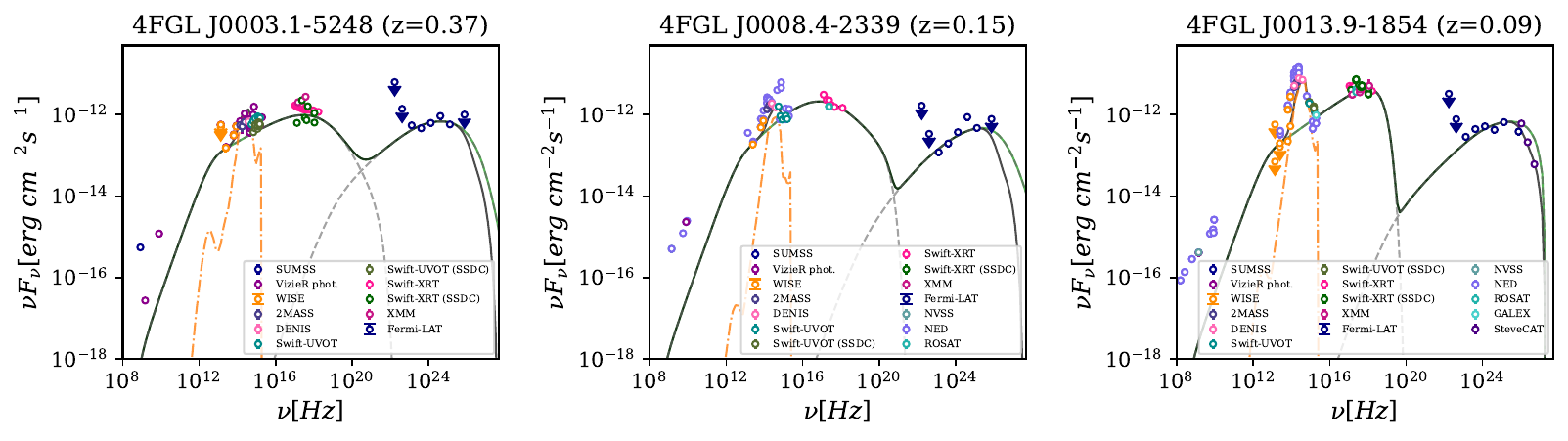} 
    \end{subfigure} 
    \hfill 
    \vspace{0.1em} 
    \begin{subfigure}[b]{\textwidth} 
        \centering 
        \includegraphics[width=0.9\textwidth]{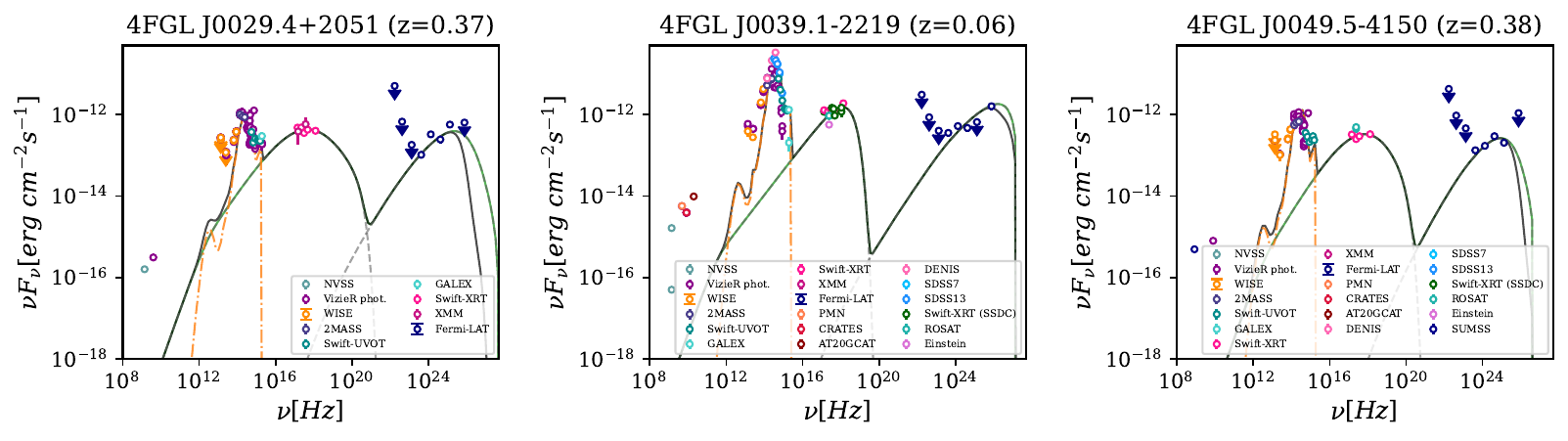} 
    \end{subfigure} 
    \hfill 
    \vspace{0.1em} 
    \begin{subfigure}[b]{\textwidth} 
        \centering 
        \includegraphics[width=0.9\textwidth]{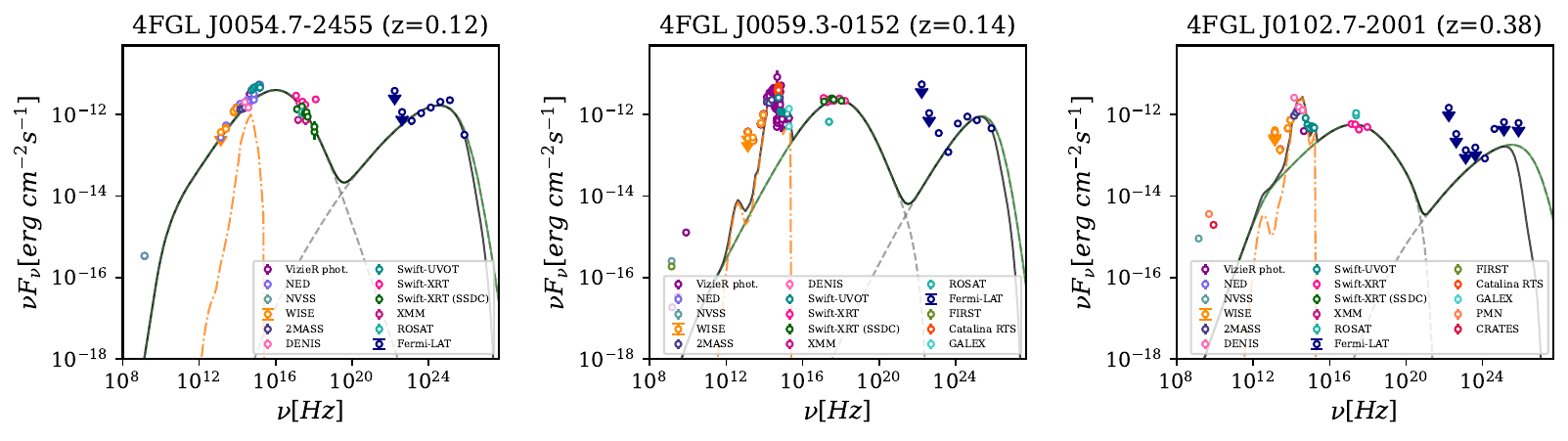} 
    \end{subfigure} 
    \hfill 
\caption{The multi-wavelength SEDs of the 124 selected sources are presented with their best-fit models. The fluxes used to reconstruct the SEDs are shown as coloured points. The synchrotron and SSC emission from the best-fit model is represented by a dashed gray line, while the host galaxy emission (based on the best-fit host galaxy model given in Table \ref{table:bestfit} for each source) is shown in orange. The total intrinsic emission, summing all components, is represented by a solid green line, and the observed emission, after accounting for EBL absorption, is shown with a solid black line. These SEDs are reconstructed using archival data from the SSDC SED builder service, along with data from \textit{Swift}-XRT, \textit{Swift}-UVOT, \textit{Fermi}-LAT, and the SteveCAT catalogue.} 
\vspace{8em}
\label{fig:mwl_seds} 
\end{figure} 

    \vspace{0.1em} 
\begin{figure}[H] 
\ContinuedFloat 
\centering 
    \begin{subfigure}[b]{\textwidth} 
        \centering 
        \includegraphics[width=0.9\textwidth]{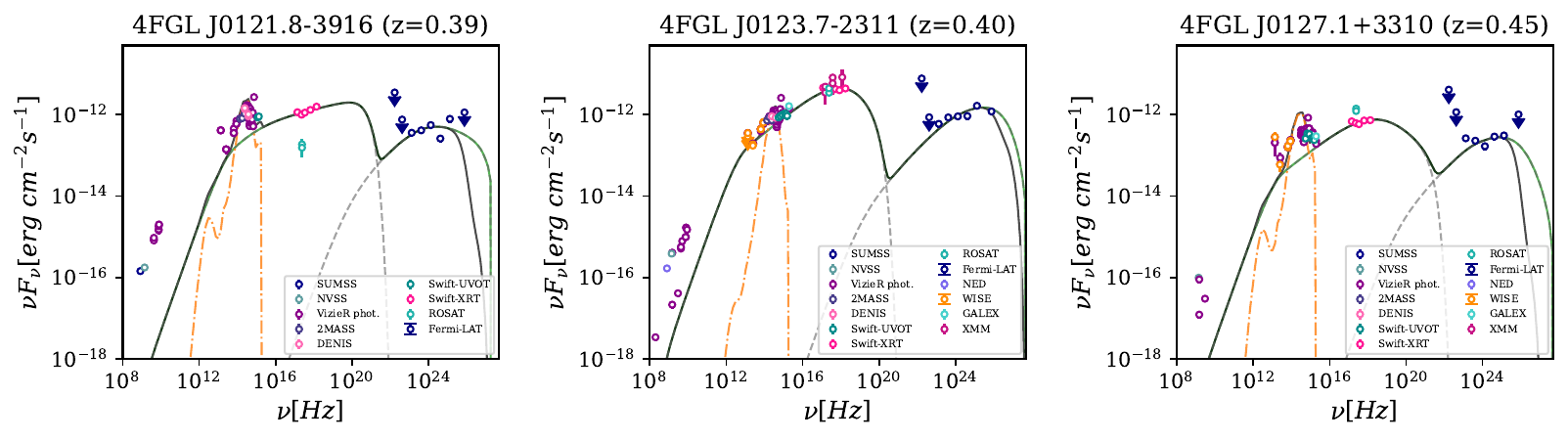} 
    \end{subfigure} 
    \hfill 
    \vspace{0.1em} 
    \begin{subfigure}[b]{\textwidth} 
        \centering 
        \includegraphics[width=0.9\textwidth]{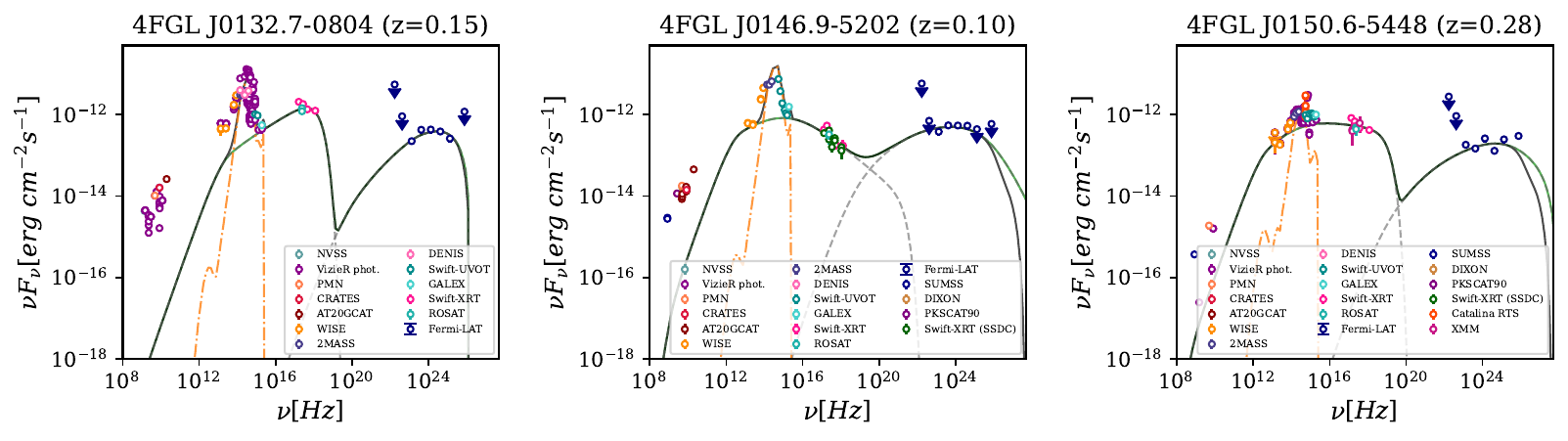} 
    \end{subfigure} 
    \hfill 
    \vspace{0.1em} 
    \begin{subfigure}[b]{\textwidth} 
        \centering 
        \includegraphics[width=0.9\textwidth]{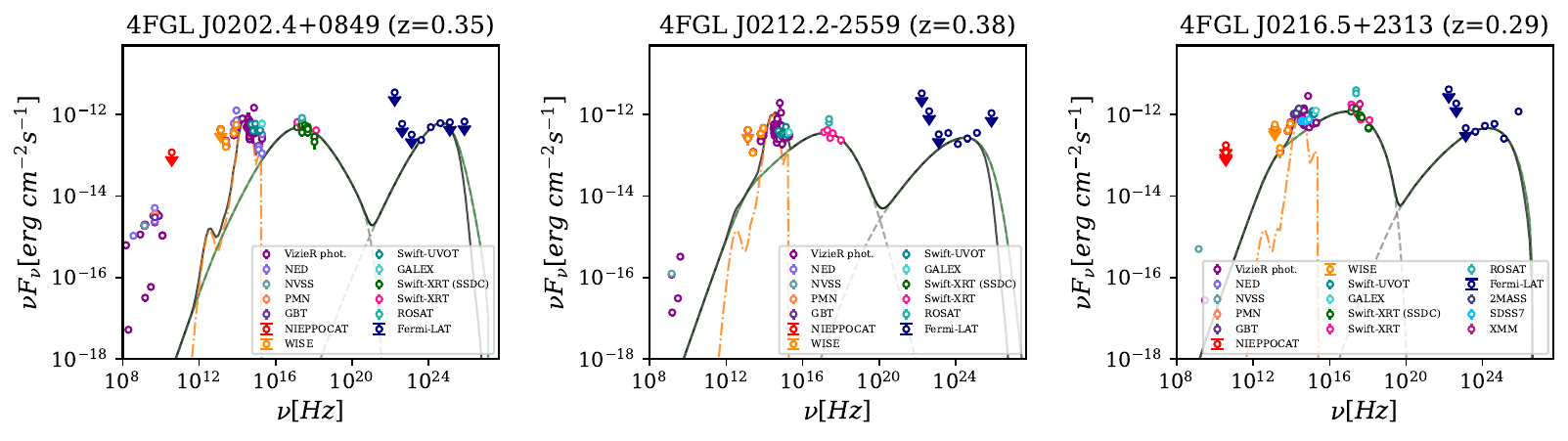} 
    \end{subfigure} 
    \hfill 
    \vspace{0.1em} 
    \begin{subfigure}[b]{\textwidth} 
        \centering 
        \includegraphics[width=0.9\textwidth]{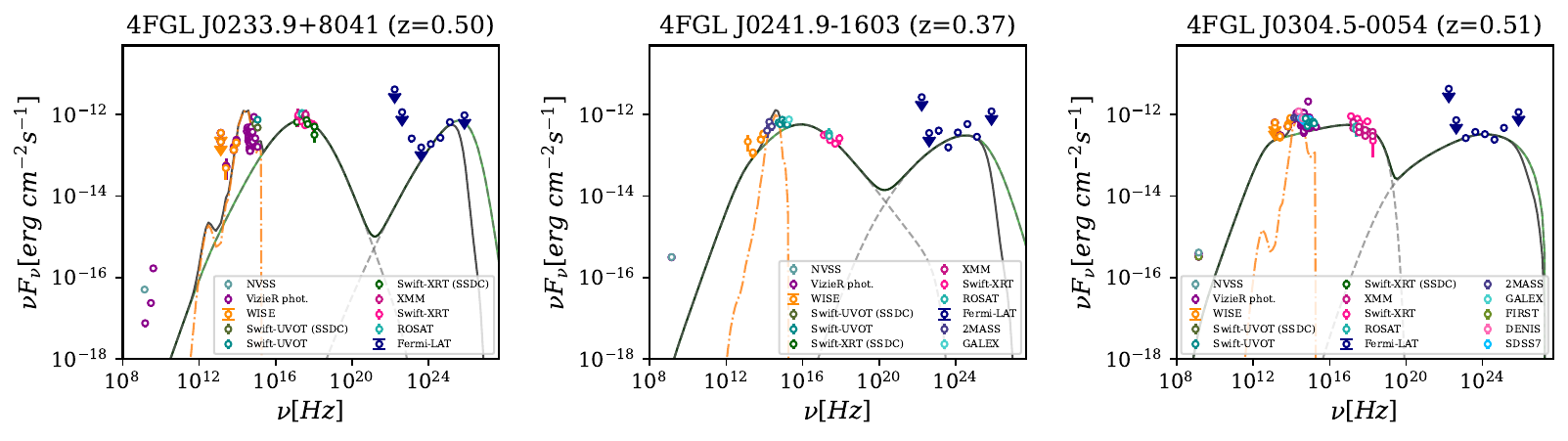} 
    \end{subfigure} 
    \hfill 
    \vspace{0.1em} 
    \begin{subfigure}[b]{\textwidth} 
        \centering 
        \includegraphics[width=0.9\textwidth]{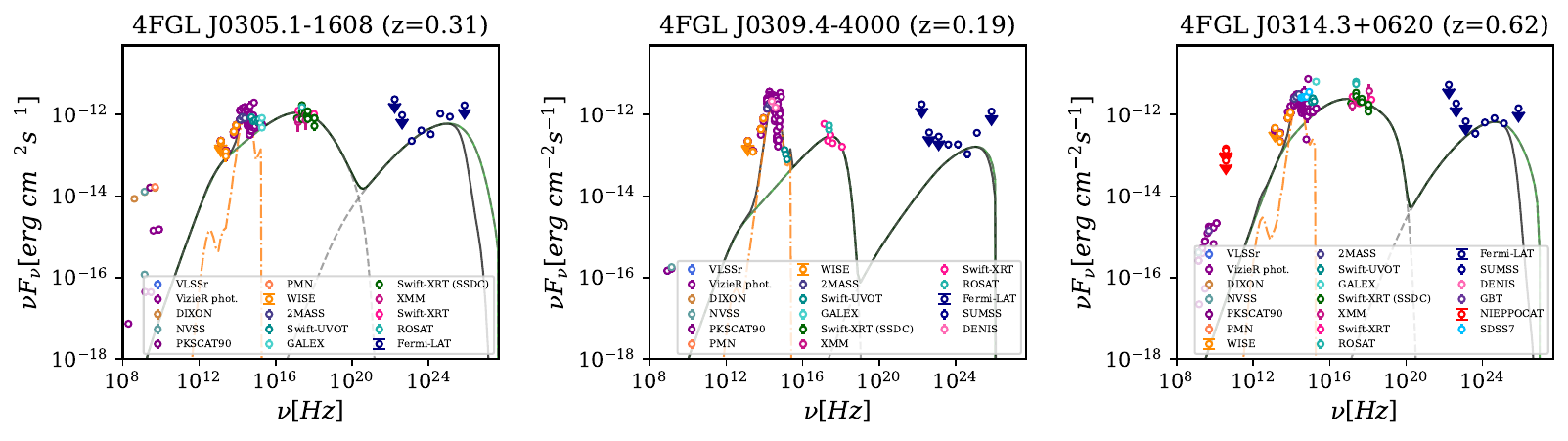} 
    \end{subfigure} 
    \hfill 
\caption{\textit{continued}} 
\end{figure} 

    \vspace{0.1em} 
\begin{figure}[H] 
\ContinuedFloat 
\centering 
    \begin{subfigure}[b]{\textwidth} 
        \centering 
        \includegraphics[width=0.9\textwidth]{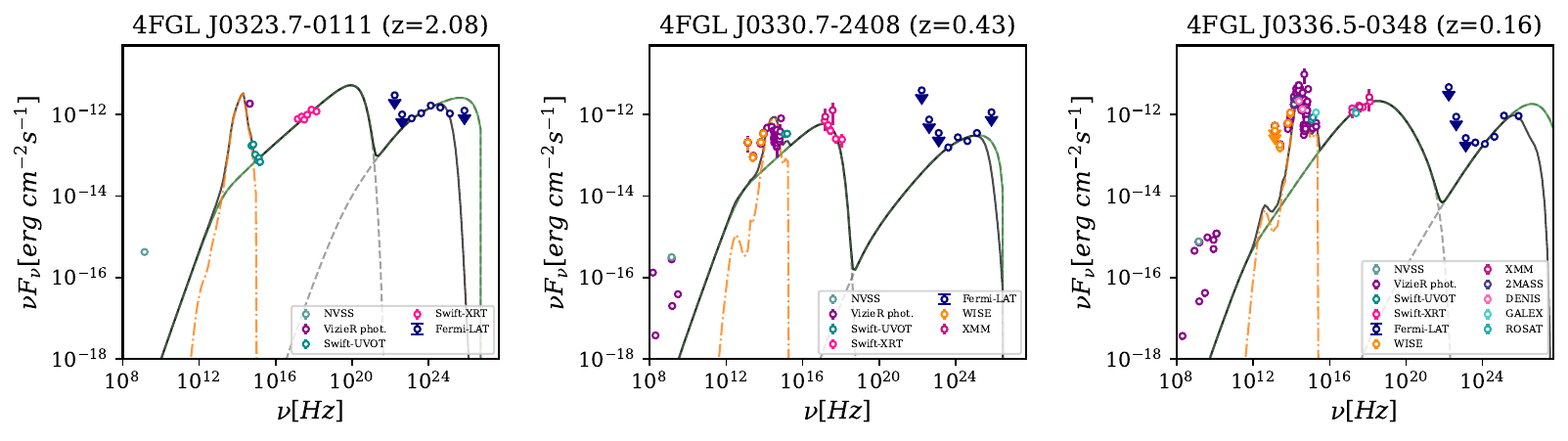} 
    \end{subfigure} 
    \hfill 
    \vspace{0.1em} 
    \begin{subfigure}[b]{\textwidth} 
        \centering 
        \includegraphics[width=0.9\textwidth]{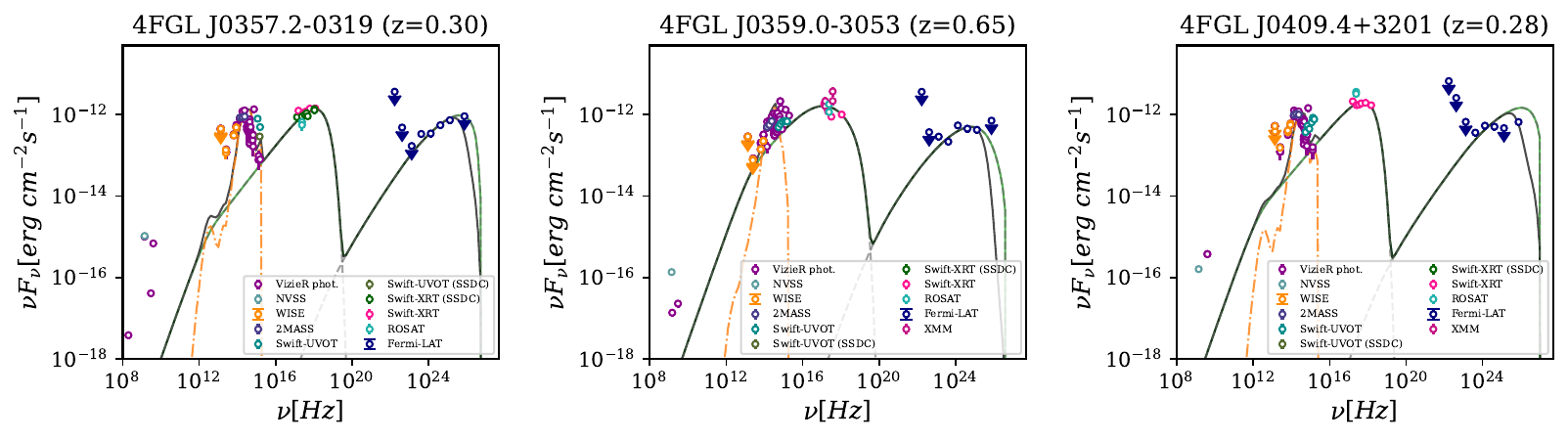} 
    \end{subfigure} 
    \hfill 
    \vspace{0.1em} 
    \begin{subfigure}[b]{\textwidth} 
        \centering 
        \includegraphics[width=0.9\textwidth]{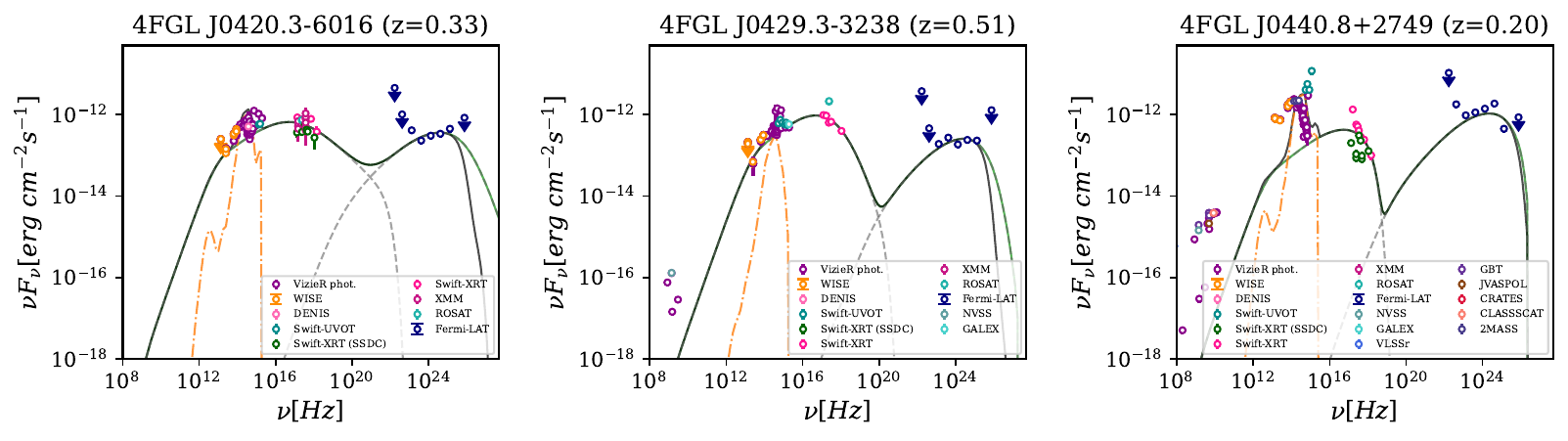} 
    \end{subfigure} 
    \hfill 
    \vspace{0.1em} 
    \begin{subfigure}[b]{\textwidth} 
        \centering 
        \includegraphics[width=0.9\textwidth]{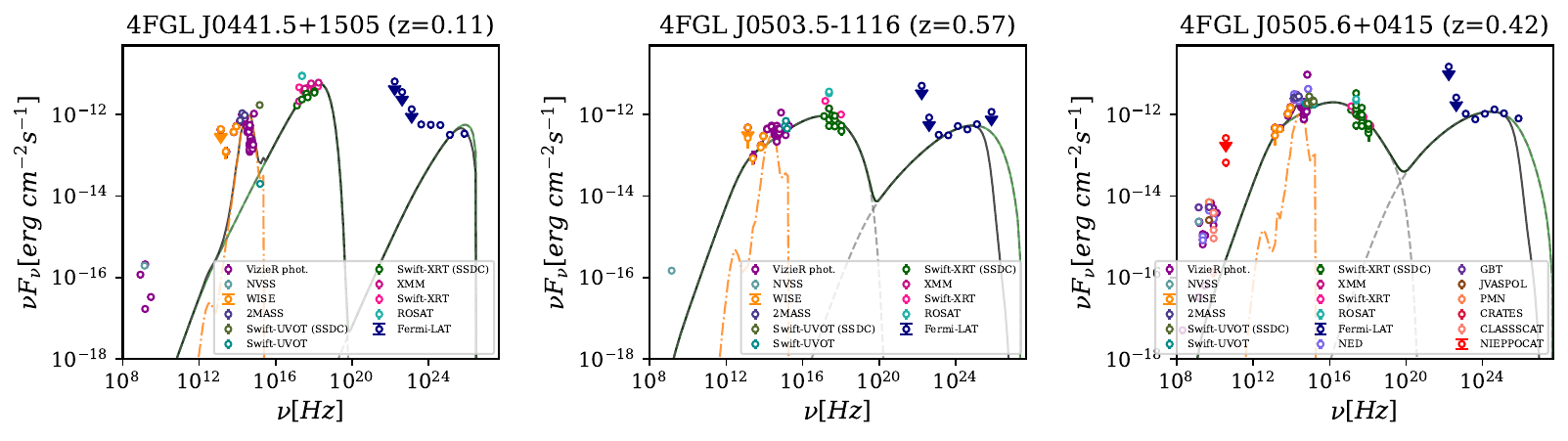} 
    \end{subfigure} 
    \hfill 
    \vspace{0.1em} 
    \begin{subfigure}[b]{\textwidth} 
        \centering 
        \includegraphics[width=0.9\textwidth]{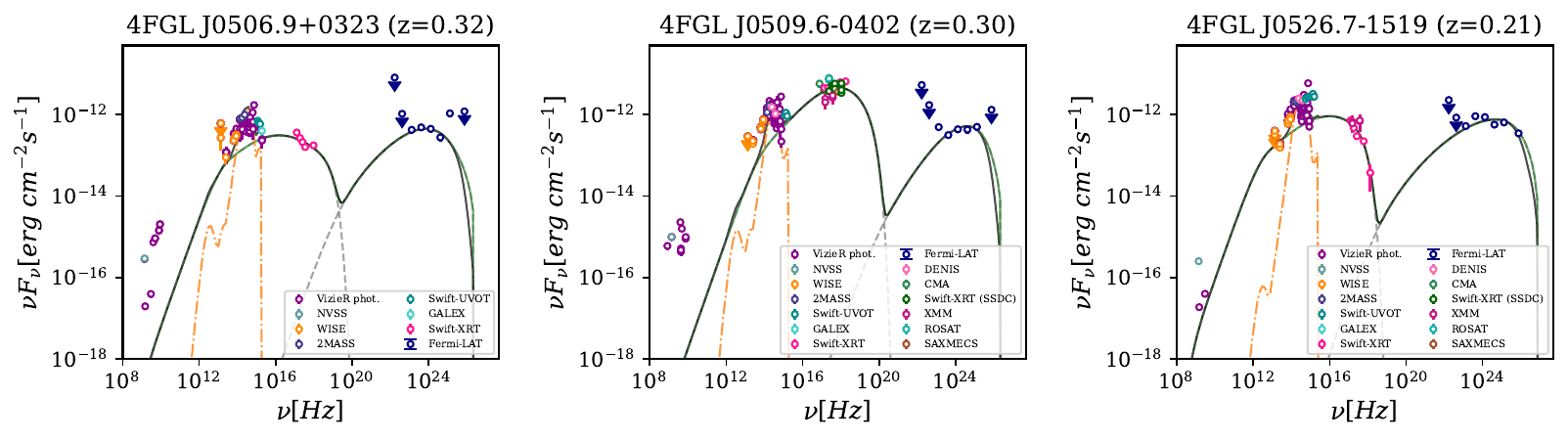} 
    \end{subfigure} 
    \hfill 
\caption{\textit{continued}} 
\end{figure} 

    \vspace{0.1em} 
\begin{figure}[H] 
\ContinuedFloat 
\centering 
    \begin{subfigure}[b]{\textwidth} 
        \centering 
        \includegraphics[width=0.9\textwidth]{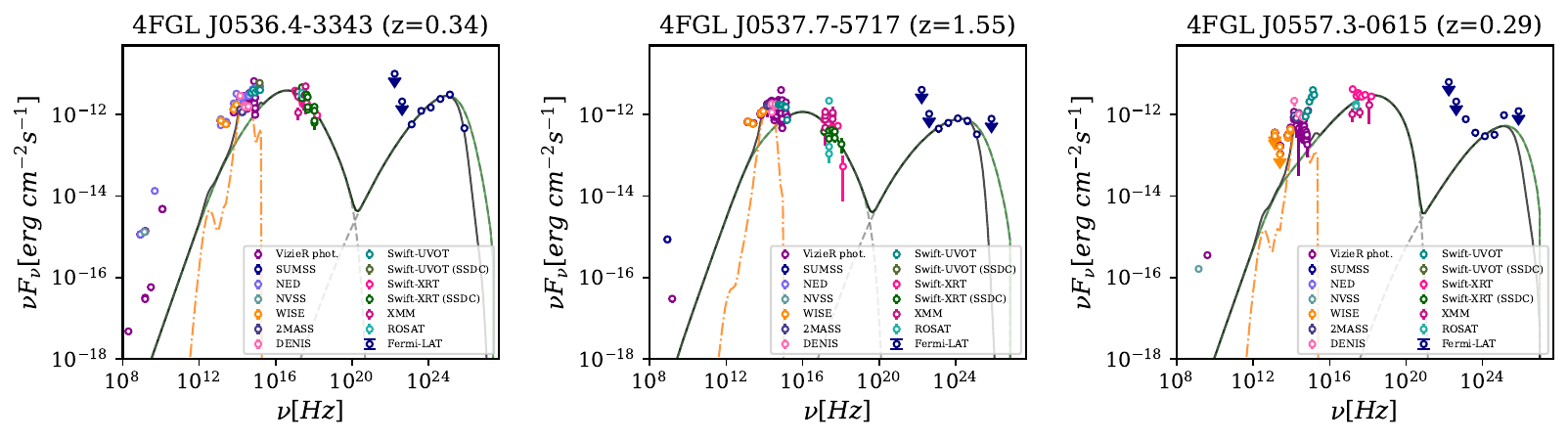} 
    \end{subfigure} 
    \hfill 
    \vspace{0.1em} 
    \begin{subfigure}[b]{\textwidth} 
        \centering 
        \includegraphics[width=0.9\textwidth]{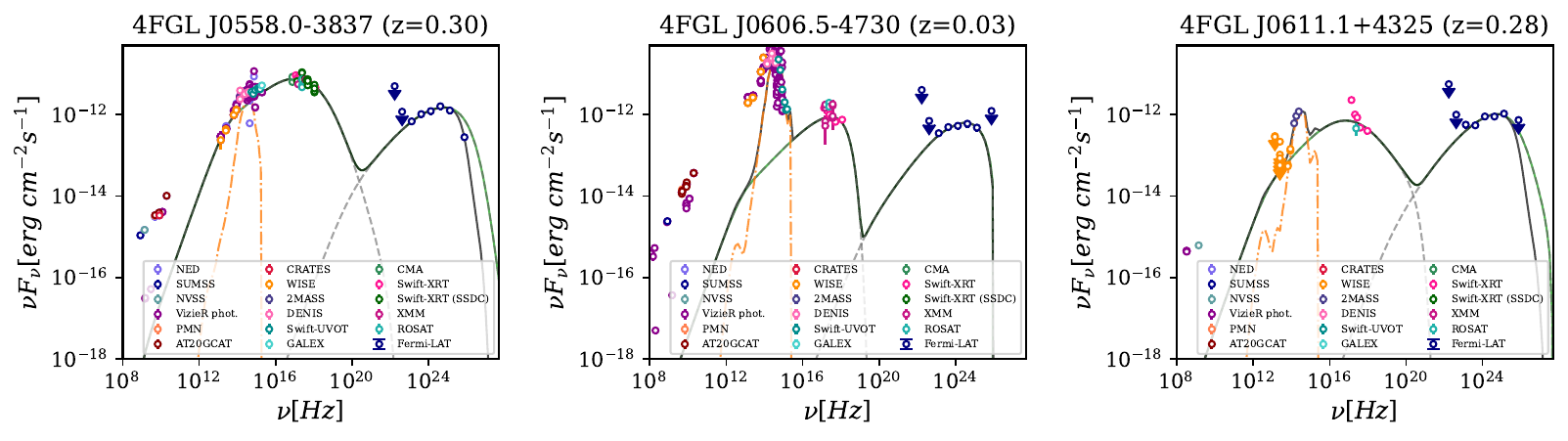} 
    \end{subfigure} 
    \hfill 
    \vspace{0.1em} 
    \begin{subfigure}[b]{\textwidth} 
        \centering 
        \includegraphics[width=0.9\textwidth]{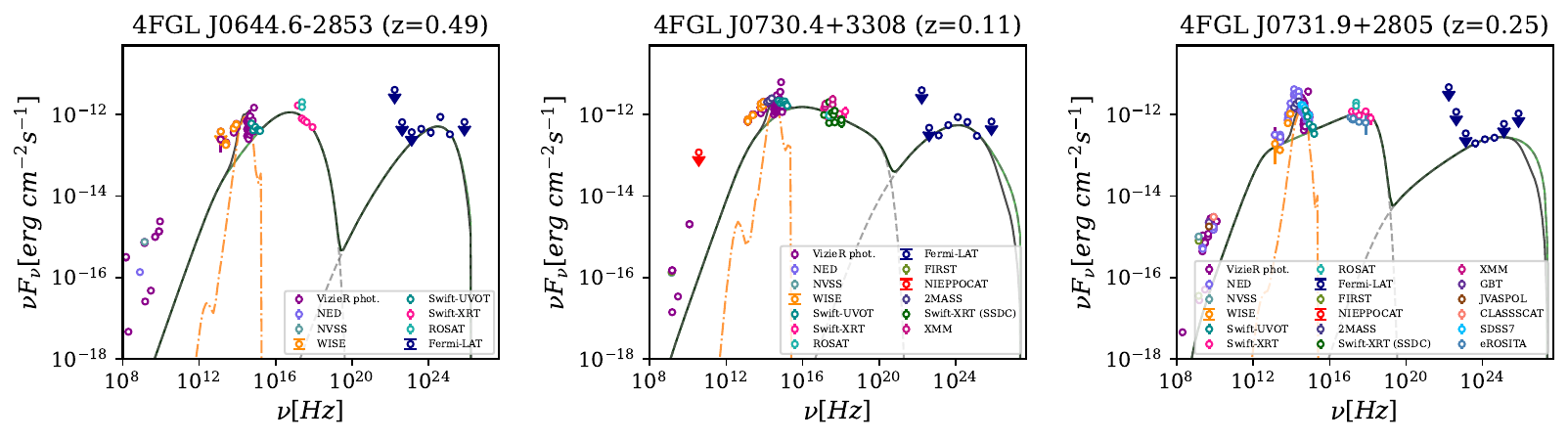} 
    \end{subfigure} 
    \hfill 
    \vspace{0.1em} 
    \begin{subfigure}[b]{\textwidth} 
        \centering 
        \includegraphics[width=0.9\textwidth]{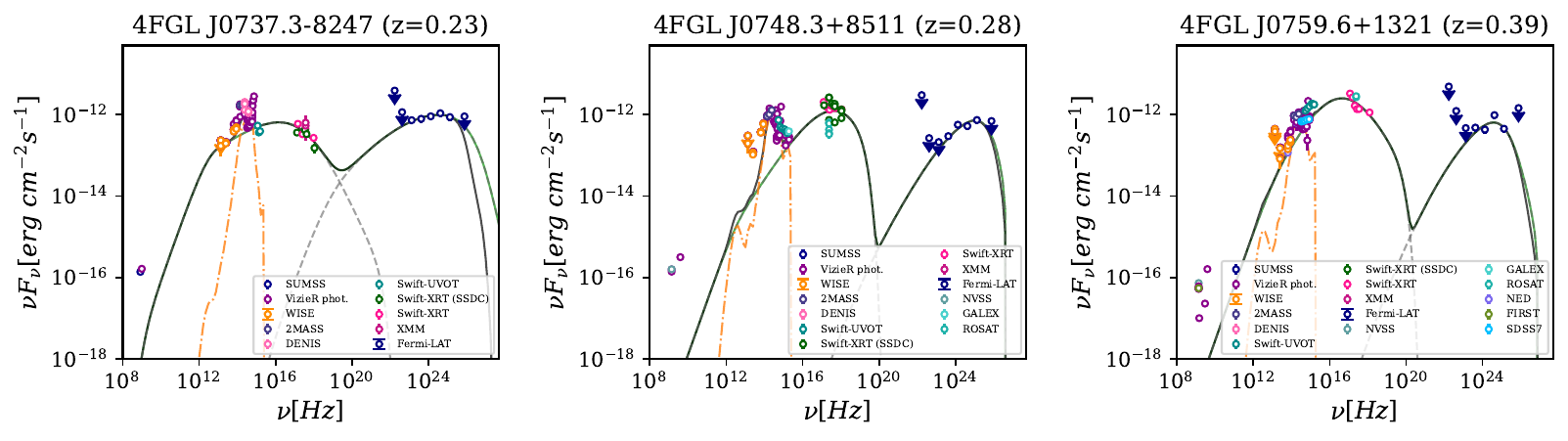} 
    \end{subfigure} 
    \hfill 
    \vspace{0.1em} 
    \begin{subfigure}[b]{\textwidth} 
        \centering 
        \includegraphics[width=0.9\textwidth]{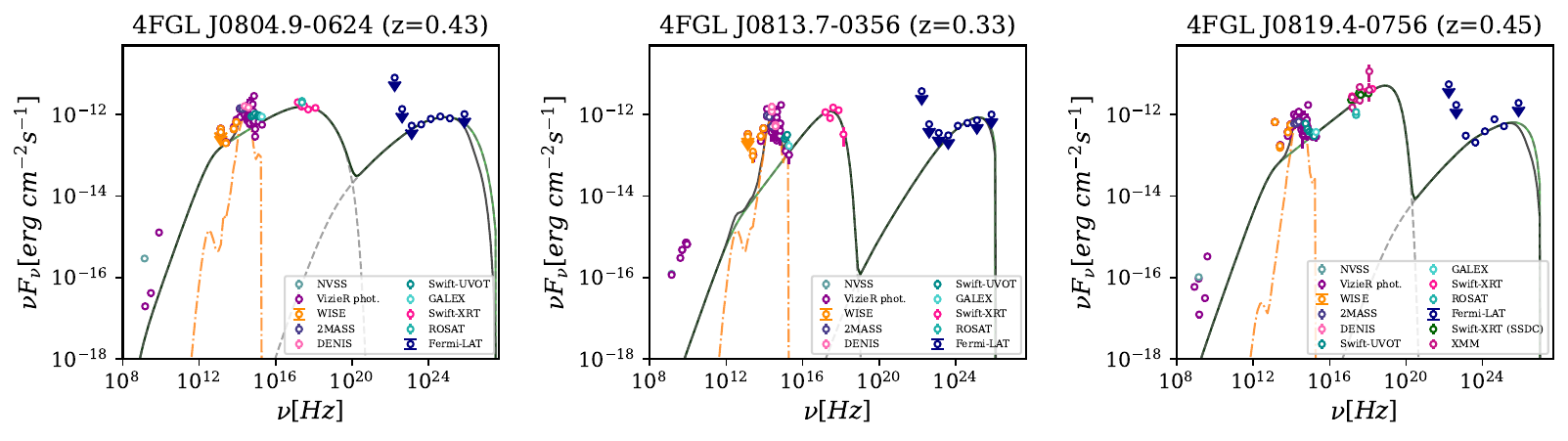} 
    \end{subfigure} 
    \hfill 
\caption{\textit{continued}} 
\end{figure} 

    \vspace{0.1em} 
\begin{figure}[H] 
\ContinuedFloat 
\centering 
    \begin{subfigure}[b]{\textwidth} 
        \centering 
        \includegraphics[width=0.9\textwidth]{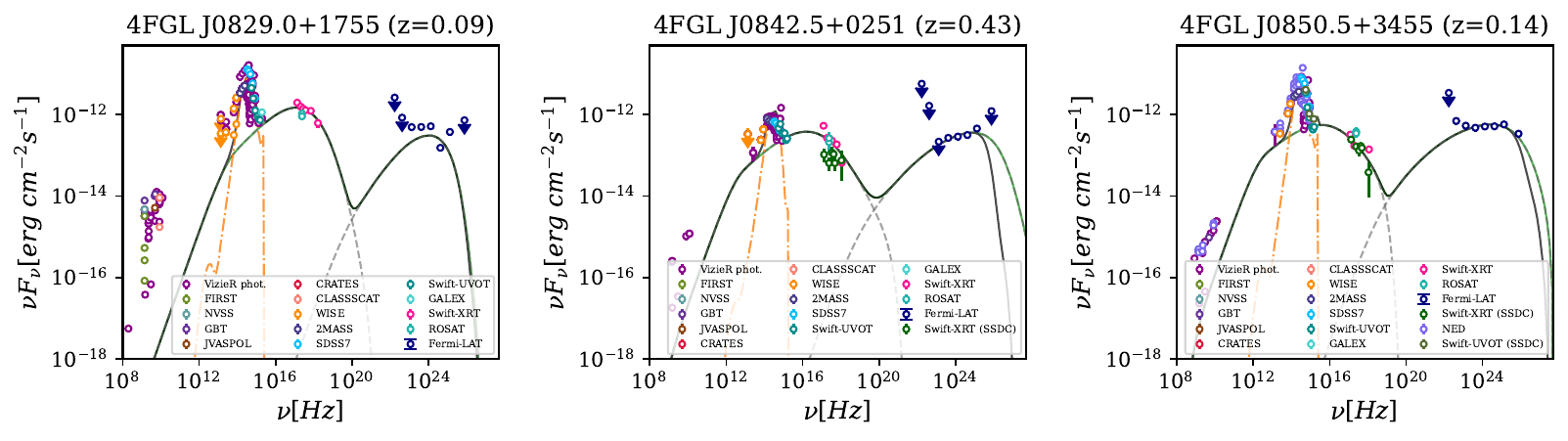} 
    \end{subfigure} 
    \hfill 
    \vspace{0.1em} 
    \begin{subfigure}[b]{\textwidth} 
        \centering 
        \includegraphics[width=0.9\textwidth]{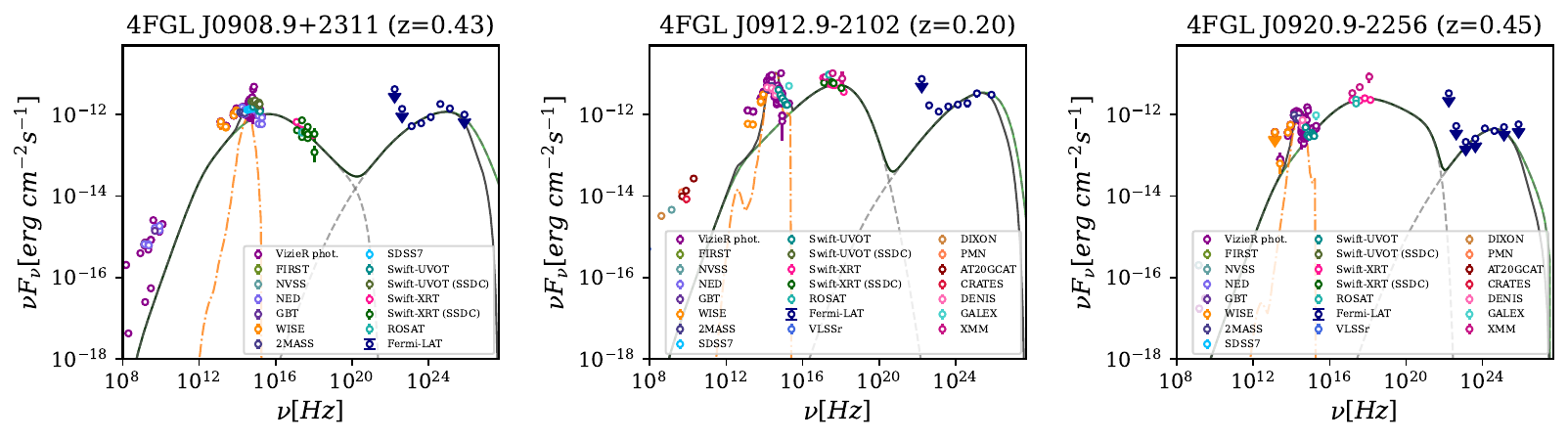} 
    \end{subfigure} 
    \hfill 
    \vspace{0.1em} 
    \begin{subfigure}[b]{\textwidth} 
        \centering 
        \includegraphics[width=0.9\textwidth]{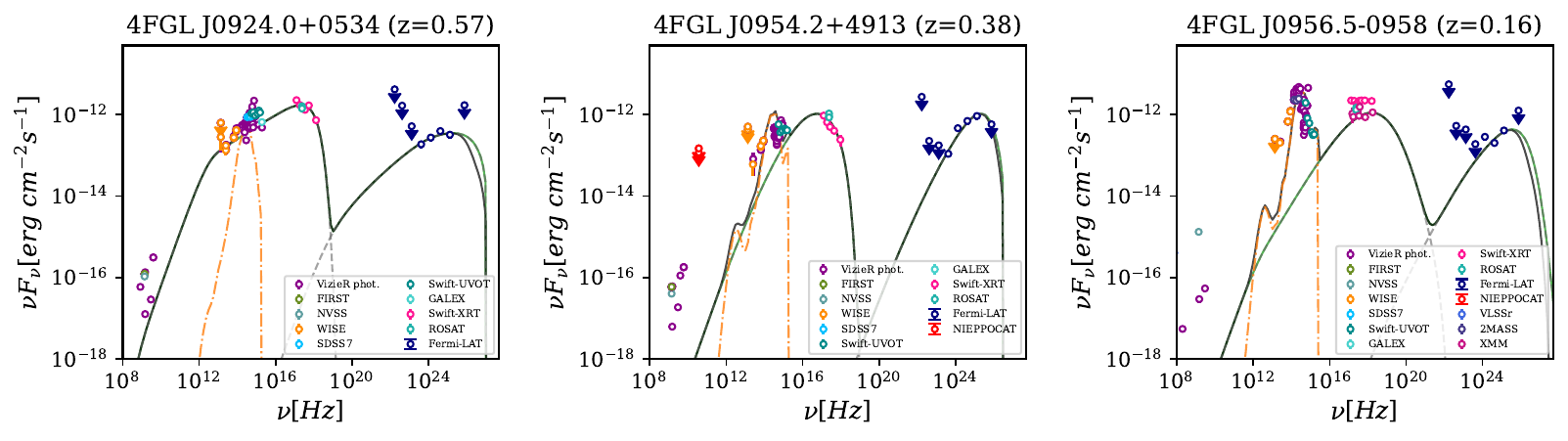} 
    \end{subfigure} 
    \hfill 
    \vspace{0.1em} 
    \begin{subfigure}[b]{\textwidth} 
        \centering 
        \includegraphics[width=0.9\textwidth]{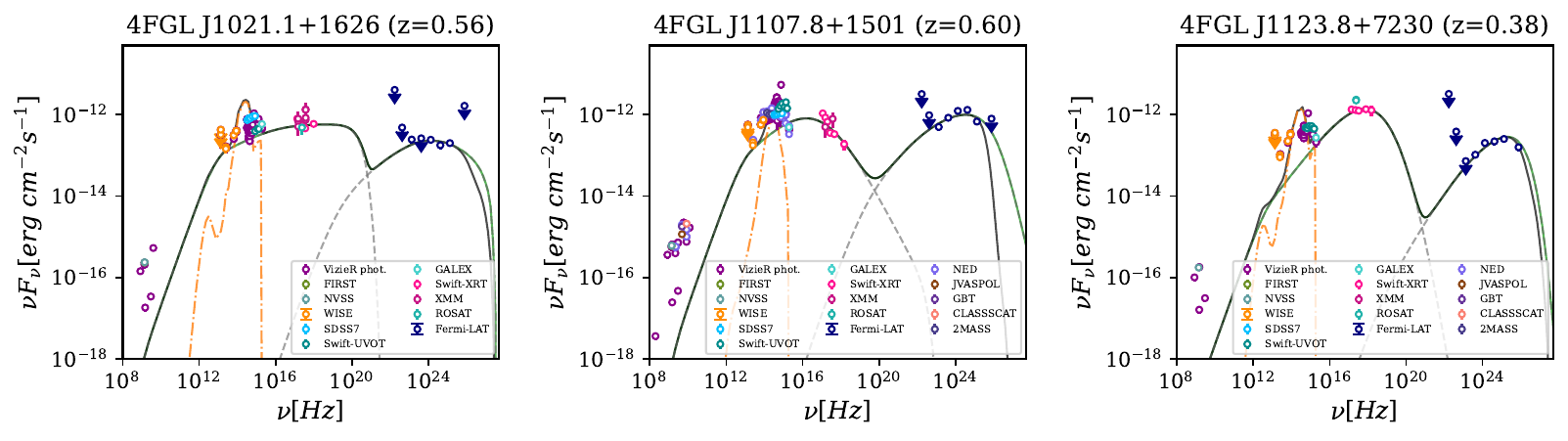} 
    \end{subfigure} 
    \hfill 
    \vspace{0.1em} 
    \begin{subfigure}[b]{\textwidth} 
        \centering 
        \includegraphics[width=0.9\textwidth]{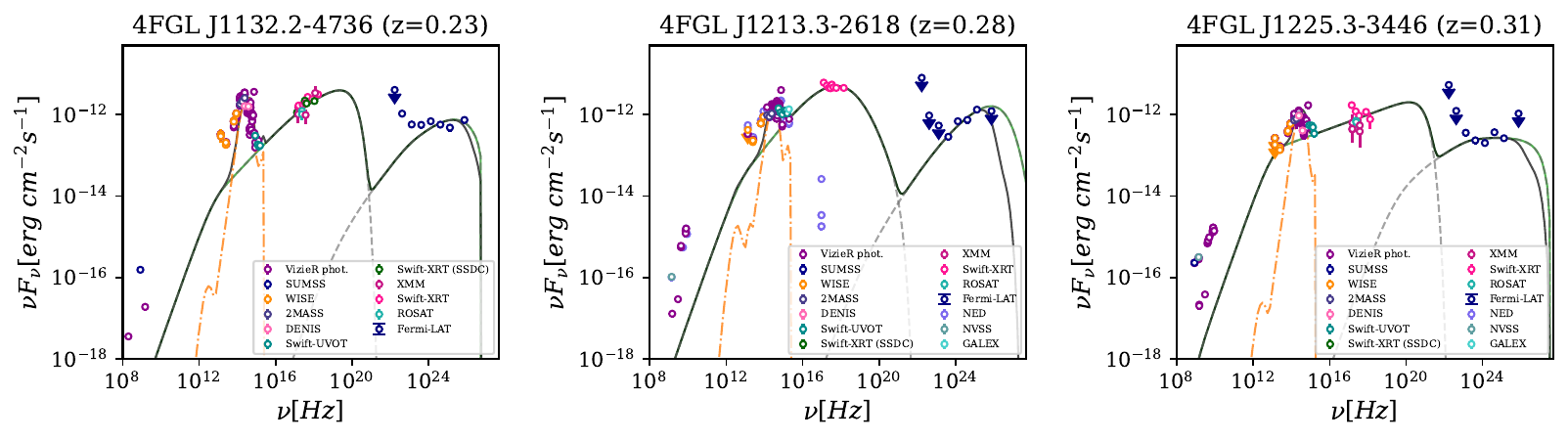} 
    \end{subfigure} 
    \hfill 
\caption{\textit{continued}} 
\end{figure} 

    \vspace{0.1em} 
\begin{figure}[H] 
\ContinuedFloat 
\centering 
    \begin{subfigure}[b]{\textwidth} 
        \centering 
        \includegraphics[width=0.9\textwidth]{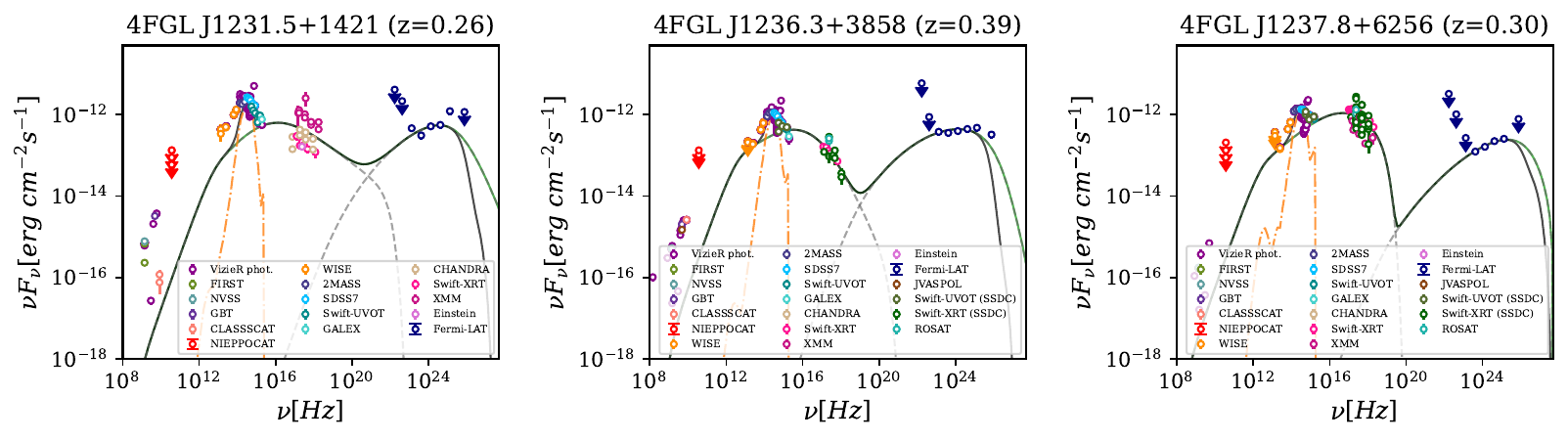} 
    \end{subfigure} 
    \hfill 
    \vspace{0.1em} 
    \begin{subfigure}[b]{\textwidth} 
        \centering 
        \includegraphics[width=0.9\textwidth]{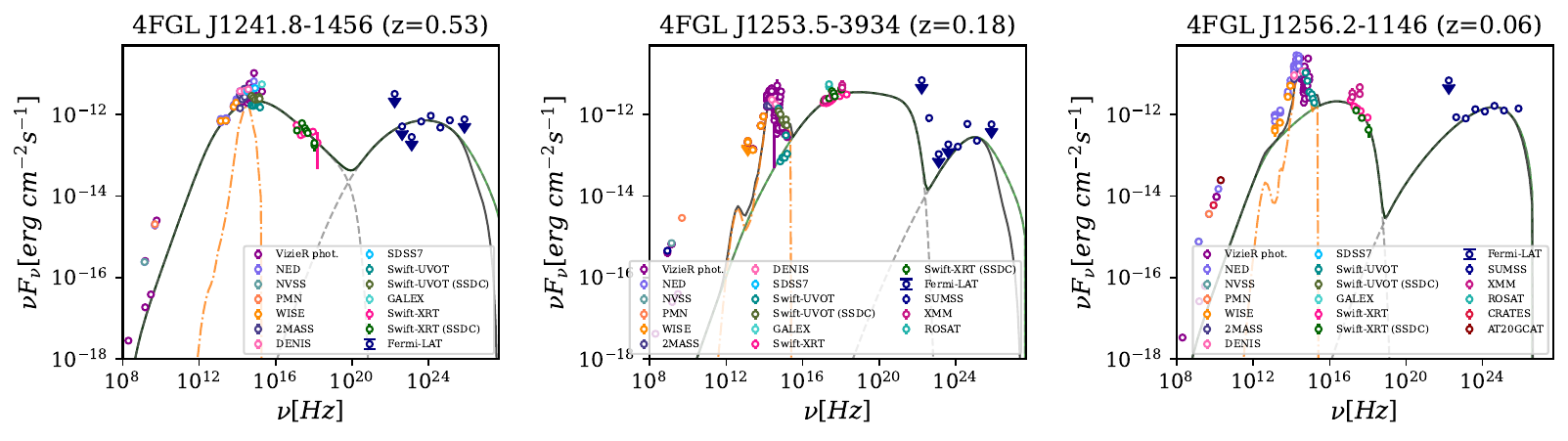} 
    \end{subfigure} 
    \hfill 
    \vspace{0.1em} 
    \begin{subfigure}[b]{\textwidth} 
        \centering 
        \includegraphics[width=0.9\textwidth]{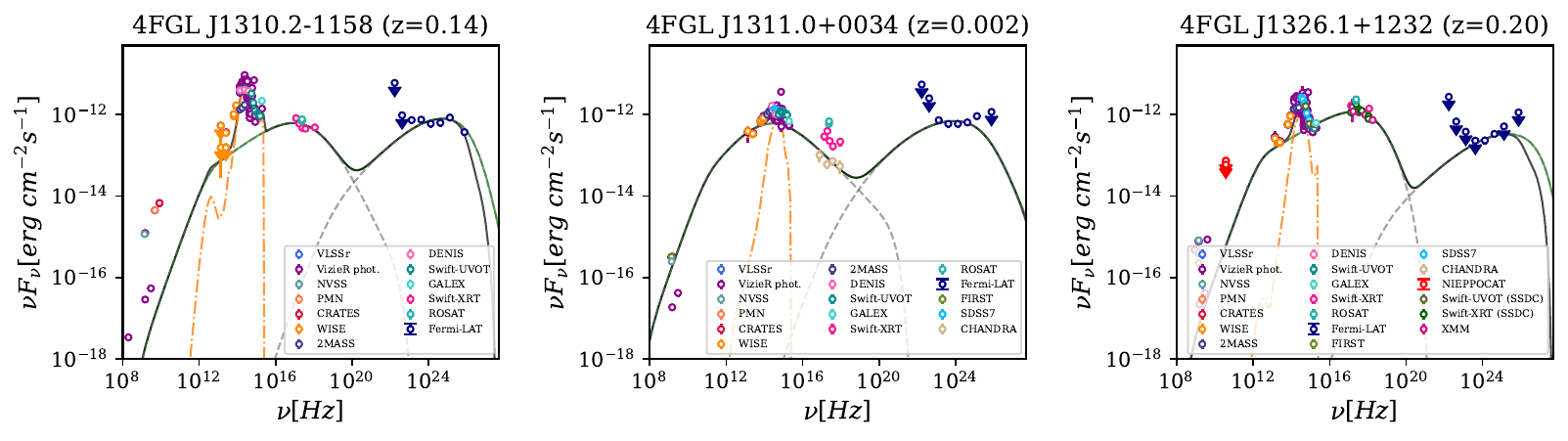} 
    \end{subfigure} 
    \hfill 
    \vspace{0.1em} 
    \begin{subfigure}[b]{\textwidth} 
        \centering 
        \includegraphics[width=0.9\textwidth]{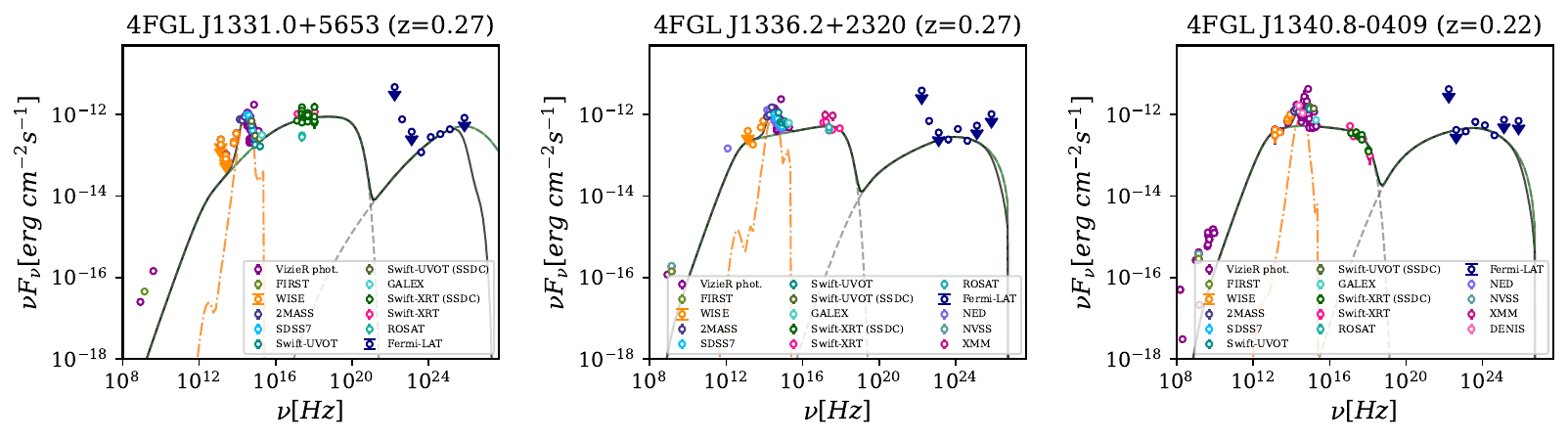} 
    \end{subfigure} 
    \hfill 
    \vspace{0.1em} 
    \begin{subfigure}[b]{\textwidth} 
        \centering 
        \includegraphics[width=0.9\textwidth]{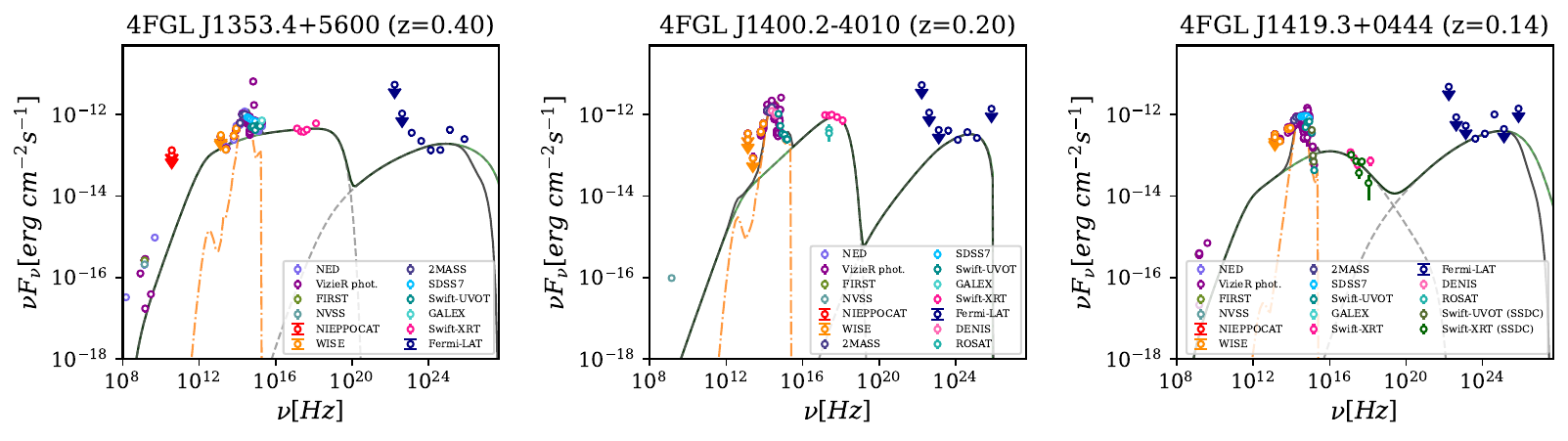} 
    \end{subfigure} 
    \hfill 
\caption{\textit{continued}} 
\end{figure} 

    \vspace{0.1em} 
\begin{figure}[H] 
\ContinuedFloat 
\centering 
    \begin{subfigure}[b]{\textwidth} 
        \centering 
        \includegraphics[width=0.9\textwidth]{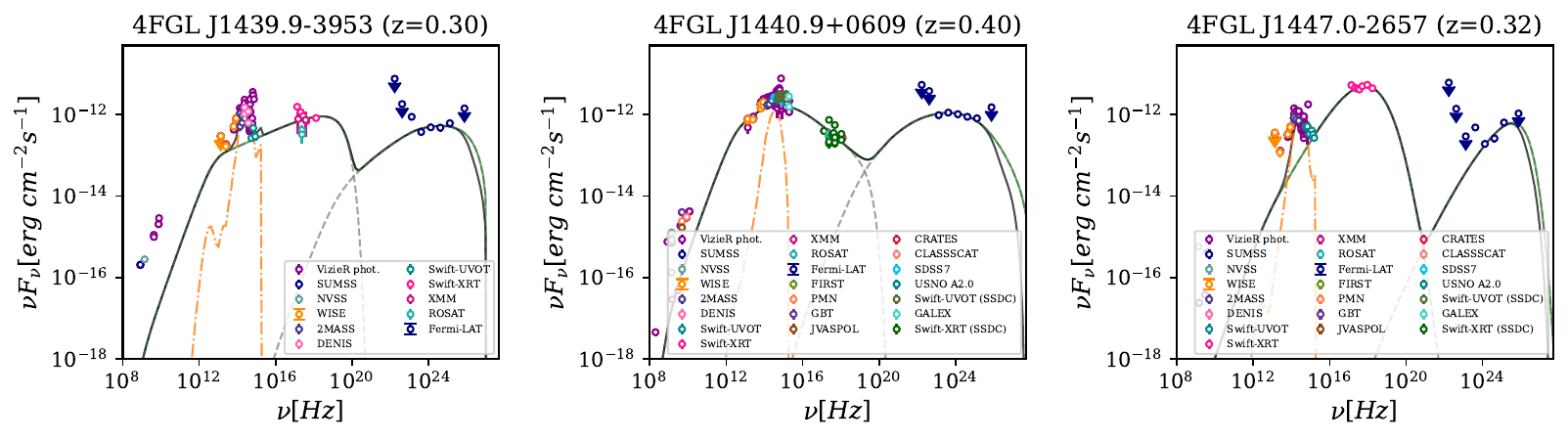} 
    \end{subfigure} 
    \hfill 
    \vspace{0.1em} 
    \begin{subfigure}[b]{\textwidth} 
        \centering 
        \includegraphics[width=0.9\textwidth]{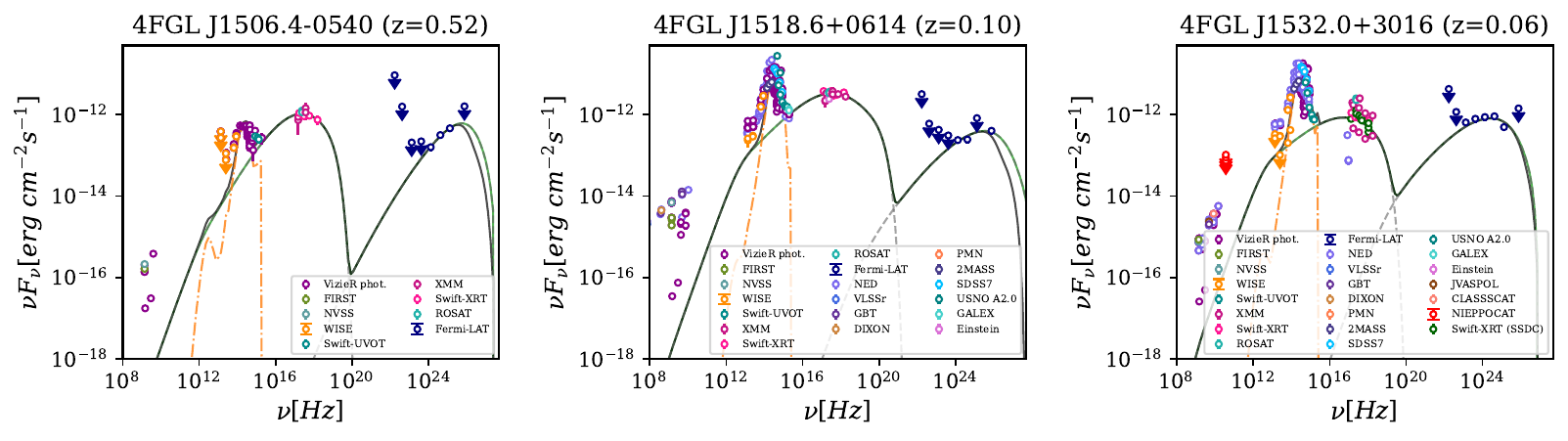} 
    \end{subfigure} 
    \hfill 
    \vspace{0.1em} 
    \begin{subfigure}[b]{\textwidth} 
        \centering 
        \includegraphics[width=0.9\textwidth]{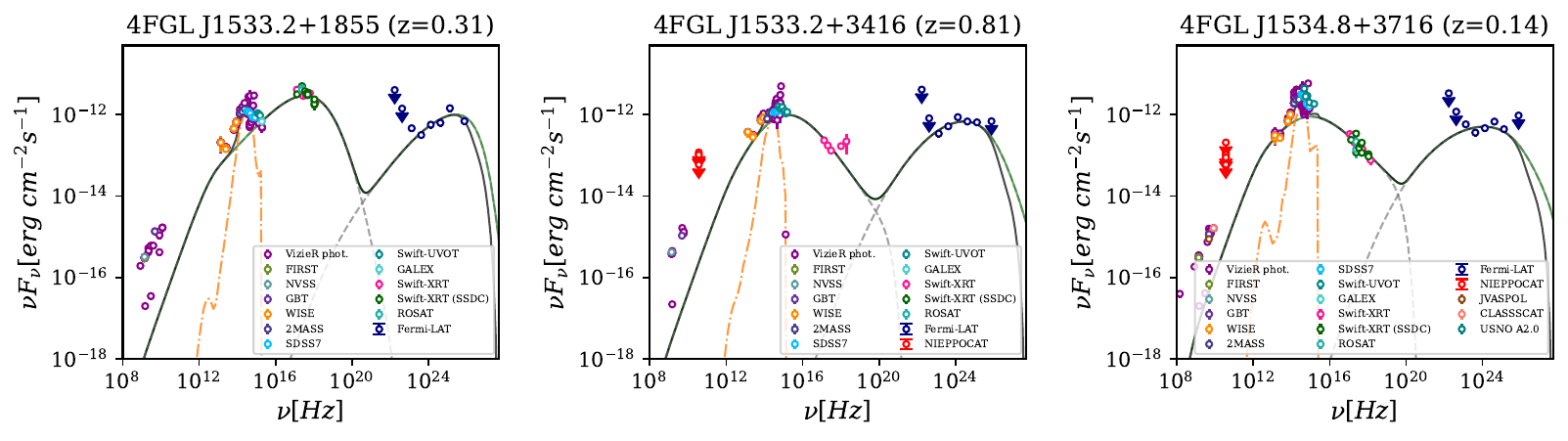} 
    \end{subfigure} 
    \hfill 
    \vspace{0.1em} 
    \begin{subfigure}[b]{\textwidth} 
        \centering 
        \includegraphics[width=0.9\textwidth]{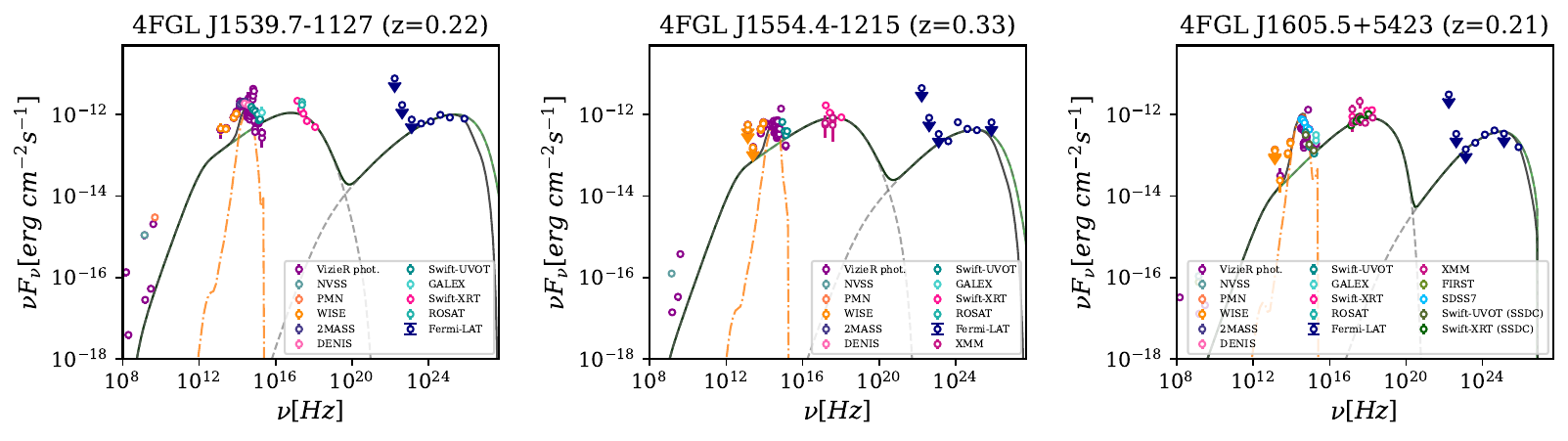} 
    \end{subfigure} 
    \hfill 
    \vspace{0.1em} 
    \begin{subfigure}[b]{\textwidth} 
        \centering 
        \includegraphics[width=0.9\textwidth]{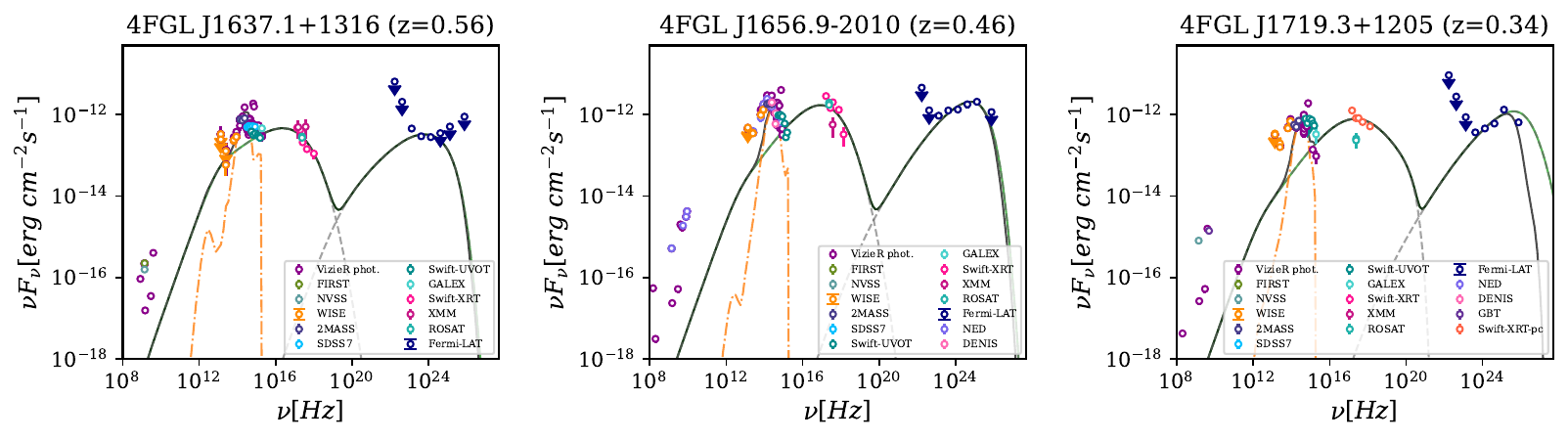} 
    \end{subfigure} 
    \hfill 
\caption{\textit{continued}} 
\end{figure} 

    \vspace{0.1em} 
\begin{figure}[H] 
\ContinuedFloat 
\centering 
    \begin{subfigure}[b]{\textwidth} 
        \centering 
        \includegraphics[width=0.9\textwidth]{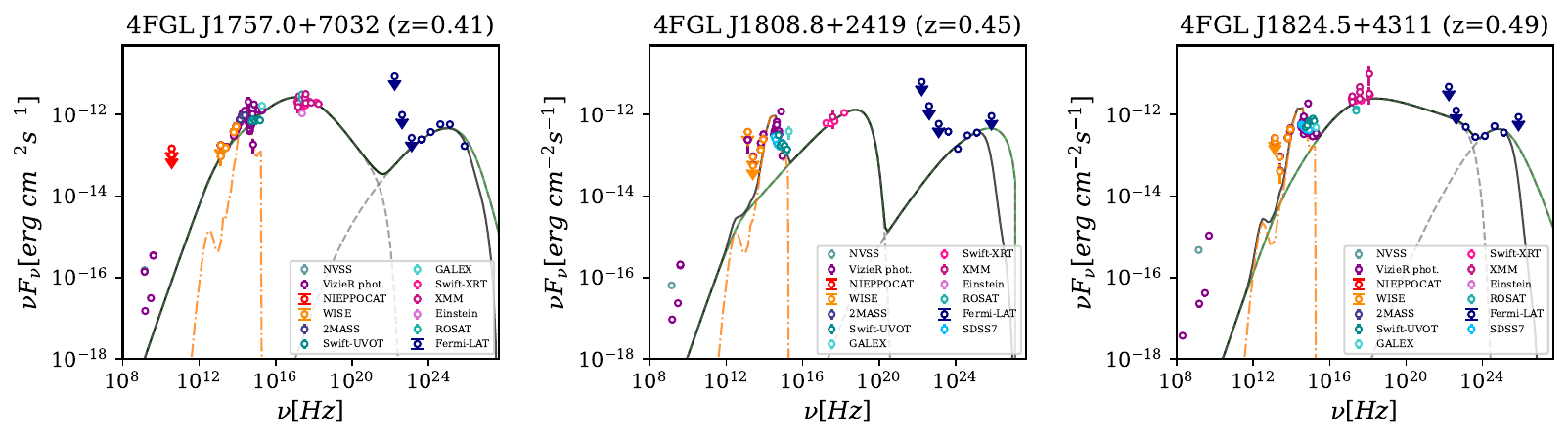} 
    \end{subfigure} 
    \hfill 
    \vspace{0.1em} 
    \begin{subfigure}[b]{\textwidth} 
        \centering 
        \includegraphics[width=0.9\textwidth]{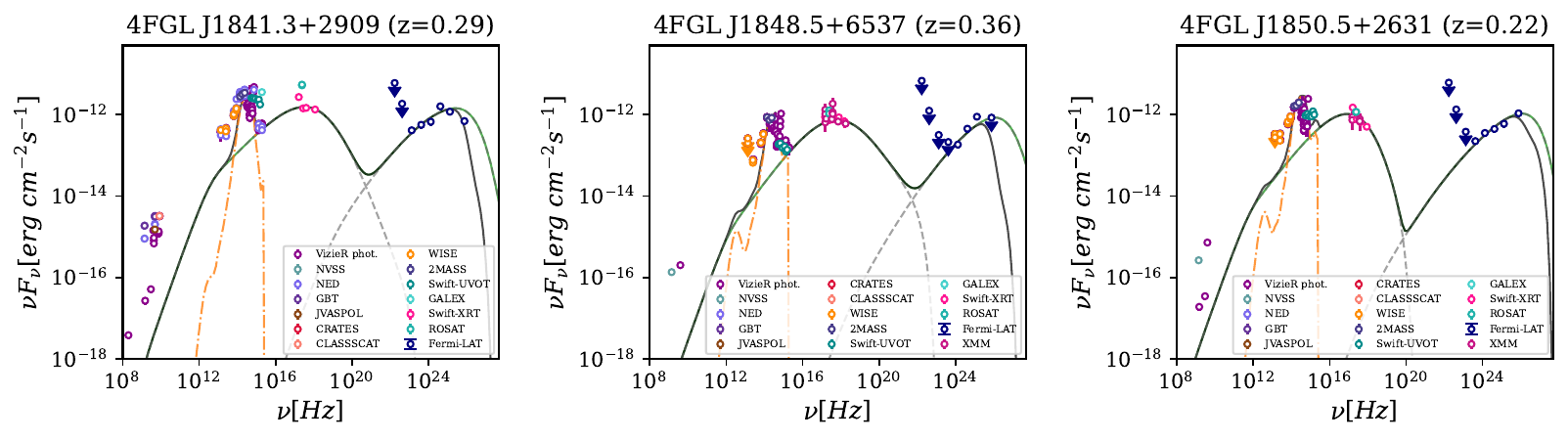} 
    \end{subfigure} 
    \hfill 
    \vspace{0.1em} 
    \begin{subfigure}[b]{\textwidth} 
        \centering 
        \includegraphics[width=0.9\textwidth]{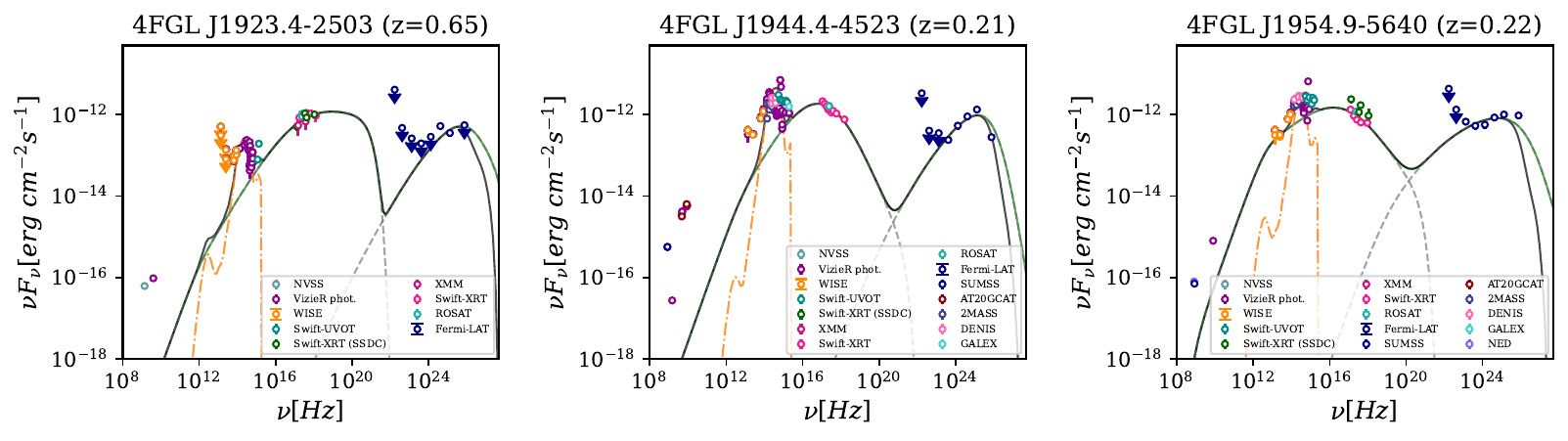} 
    \end{subfigure} 
    \hfill 
    \vspace{0.1em} 
    \begin{subfigure}[b]{\textwidth} 
        \centering 
        \includegraphics[width=0.9\textwidth]{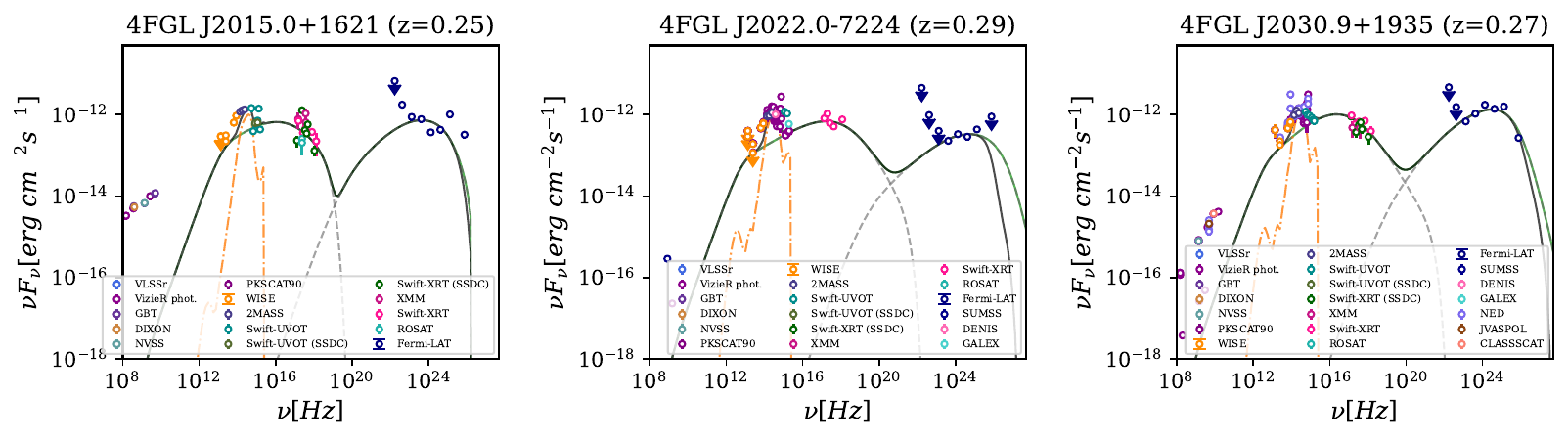} 
    \end{subfigure} 
    \hfill 
    \vspace{0.1em} 
    \begin{subfigure}[b]{\textwidth} 
        \centering 
        \includegraphics[width=0.9\textwidth]{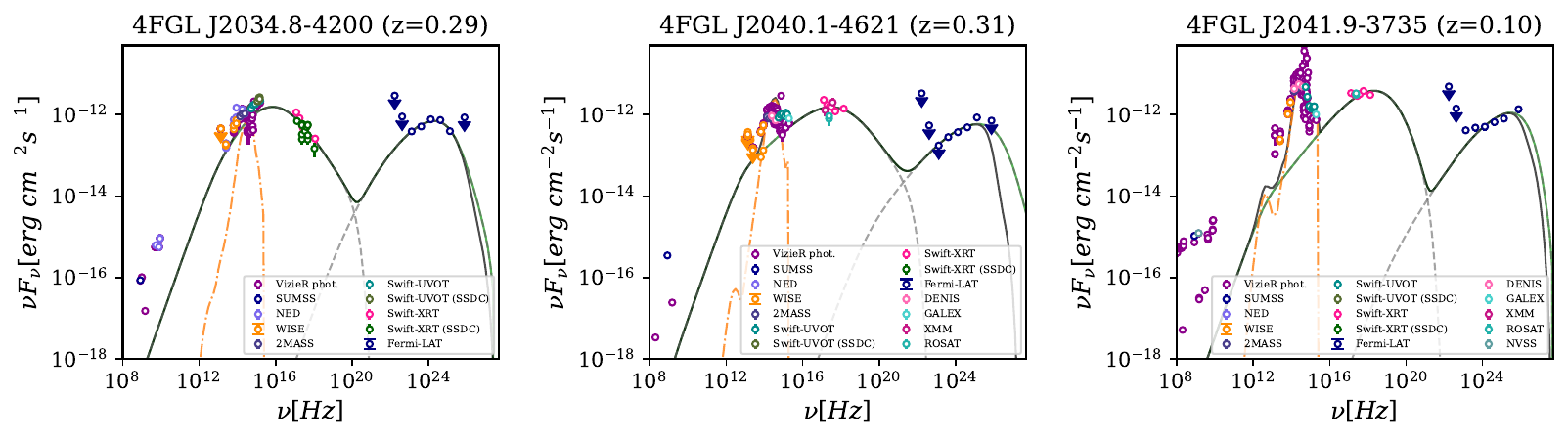} 
    \end{subfigure} 
    \hfill 
\caption{\textit{continued}} 
\end{figure} 

    \vspace{0.1em} 
\begin{figure}[H] 
\ContinuedFloat 
\centering 
    \begin{subfigure}[b]{\textwidth} 
        \centering 
        \includegraphics[width=0.9\textwidth]{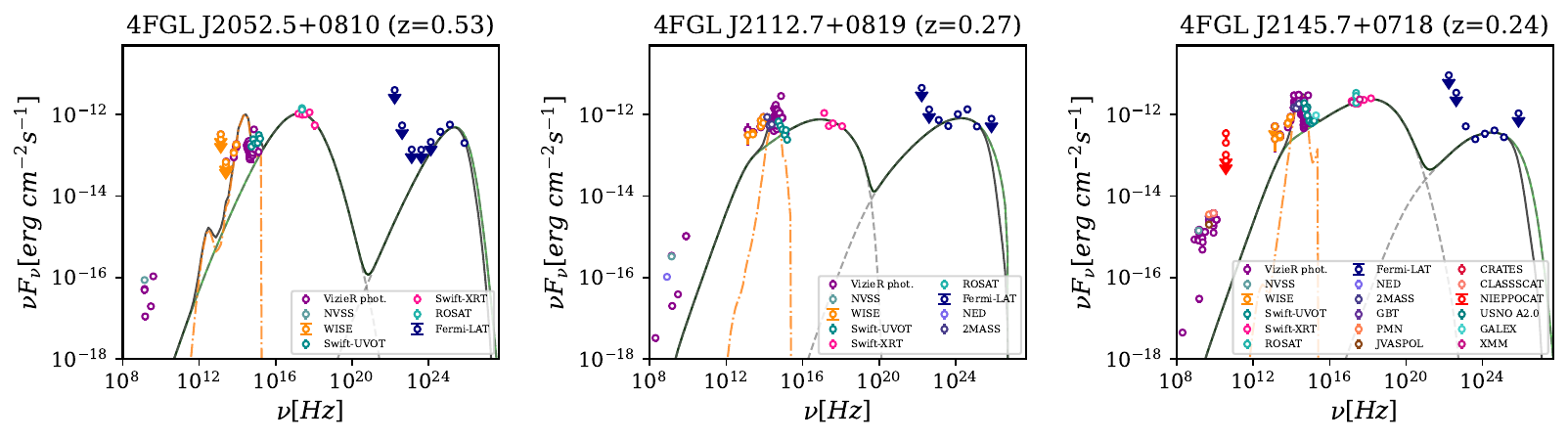} 
    \end{subfigure} 
    \hfill 
    \vspace{0.1em} 
    \begin{subfigure}[b]{\textwidth} 
        \centering 
        \includegraphics[width=0.9\textwidth]{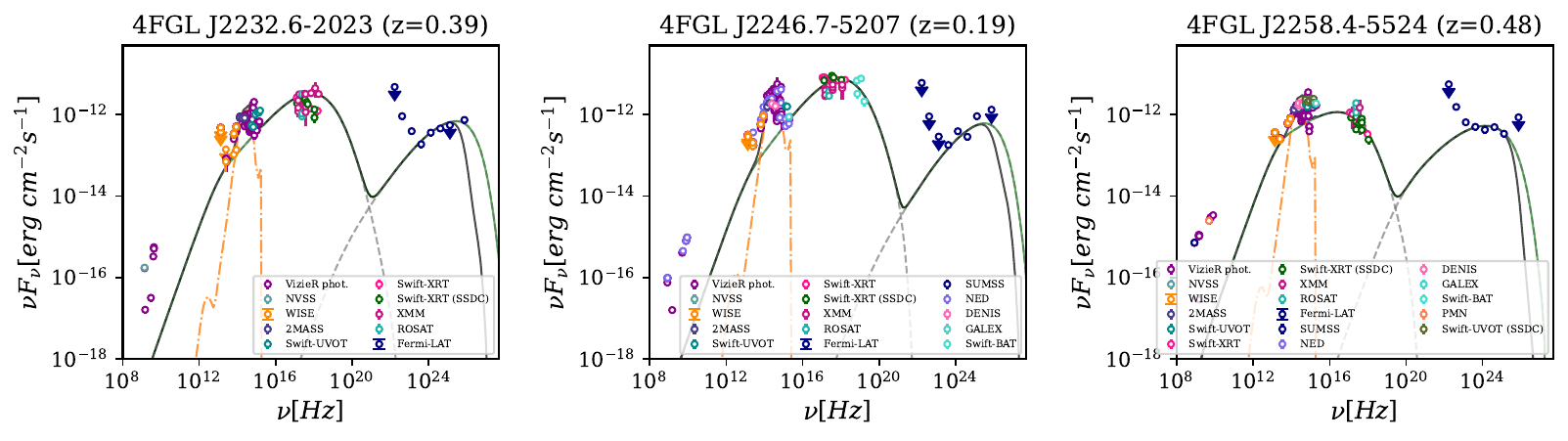} 
    \end{subfigure} 
    \hfill 
    \vspace{0.1em} 
    \begin{subfigure}[b]{\textwidth} 
        \centering 
        \includegraphics[width=0.9\textwidth]{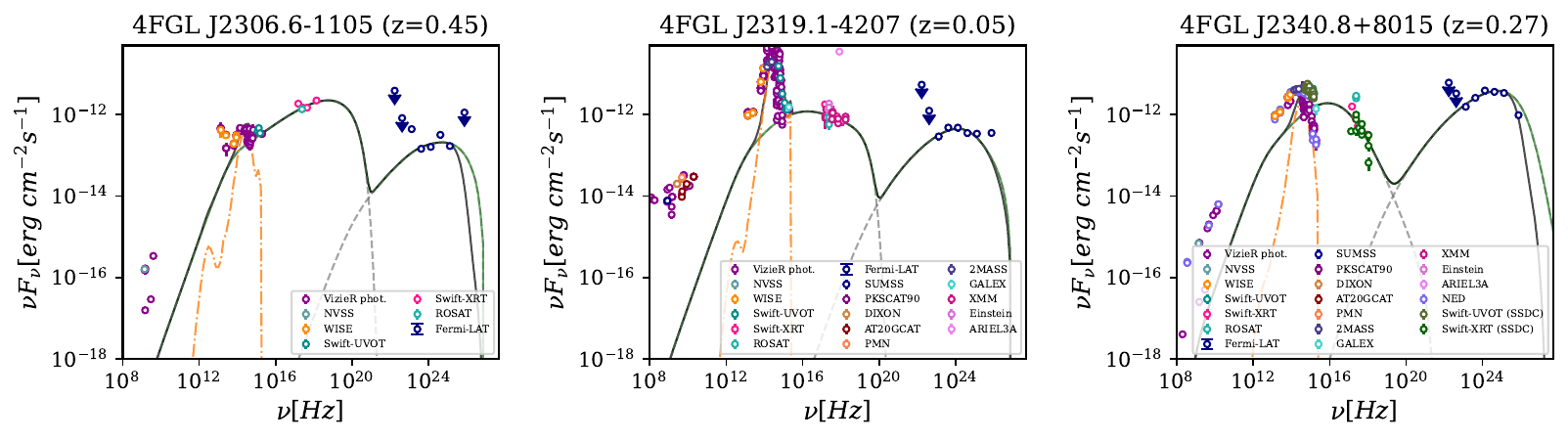} 
    \end{subfigure} 
    \hfill 
    \vspace{0.1em} 
    \begin{subfigure}[b]{\textwidth} 
        \centering 
        \includegraphics[width=0.3\textwidth]{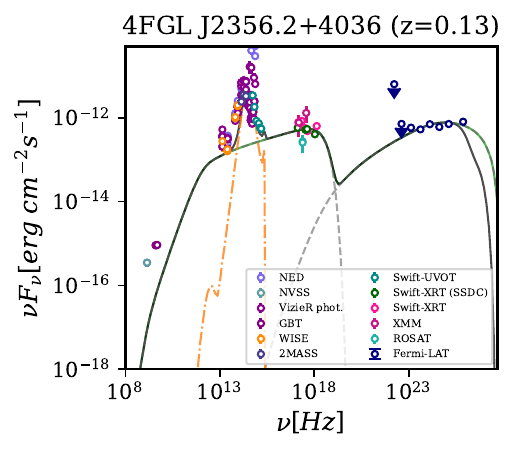} 
    \end{subfigure} 
    \hfill 
\caption{\textit{continued}} 
\end{figure}

\newpage

\section{Best-fit parameters and energetic report of the sample sources}
\label{appendix:tables}
This section provides the resulting best-fit parameters and energetic report of the sample sources. Table \ref{table:bestfit} shows the best-fit parameters obtained from the SED modelling carried out for the 124 sources of the sample using a one-zone SSC model. Table \ref{table:energy-budget} gives the $\nu_{sync}^{peak}$ and CD results, as well as the luminosity of each emitting component and the luminosity carried by the jet for the radiative components, electrons, and magnetic fields.

{\tiny 
\begin{longtable}{ccccccccccc}
\caption{{\small Best-fit parameters obtained from the broadband SED modelling, including the magnetic field strength $B$, the jet viewing angle $\theta$, the maximum Lorentz factor $\gamma_{max}$, the break Lorentz $\gamma_{break}$, the low energy and high energy spectral indices $p_1$ and $p_2$, the electron density $N$, the best-fitting host galaxy model and the resulting temperature of the host galaxy $T_{host}$, and the $\chi^2$ used to select sources with a good fitting result. The values of the radius of the emitting region $R$, the distance from the SMBH $R_H$ and the bulk Lorentz factor $\Gamma$ are fixed to $R = 10^{16}$ cm, $R_H = 2\times 10^{18}$ cm, $\Gamma = 20$.}}
\label{table:bestfit}\\
\toprule
     4FGL Name &                B [G] & \thead{$\theta $ \\ $[\deg]$} &      $\gamma_{max}$ &    $\gamma_{break}$ & $p_1$ &                $p_2$ &        N [cm$^{-3}$] & \thead{Host \\ Galaxy} & \thead{$T_{host}$ \\ $[\times 10^3 \text{K}]$} &            $\chi^2$ \\
\midrule
\endfirsthead
\caption[]{\textit{continued}} \\
\toprule
     4FGL Name &                B [G] & \thead{$\theta $ \\ $[\deg]$} &      $\gamma_{max}$ &    $\gamma_{break}$ & $p_1$ &                $p_2$ &        N [cm$^{-3}$] & \thead{Host \\ Galaxy} & \thead{$T_{host}$ \\ $[\times 10^3 \text{K}]$} &            $\chi^2$ \\
\midrule
\endhead
\midrule
\multicolumn{11}{r}{{Continued on next page}} \\
\midrule
\endfoot

\bottomrule
\endlastfoot
J0003.1$-$5248 &                 0.12 &                           5.4 & $1.8 \times 10^{7}$ & $1.3 \times 10^{6}$ &   2.4 &                  2.9 &  $2.4 \times 10^{5}$ &                     S0 &                                            4.3 &                 0.80 \\
J0008.4$-$2339 &                0.029 &                         0.017 & $6.8 \times 10^{6}$ & $1.7 \times 10^{5}$ &   2.0 &                  2.1 &  $1.0 \times 10^{3}$ &                  Ell13 &                                            4.7 &                0.68 \\
J0013.9$-$1854 &                0.063 &                           3.0 & $9.7 \times 10^{5}$ & $2.1 \times 10^{7}$ &   2.3 &                  9.7 &  $1.8 \times 10^{4}$ &                  Ell13 &                                            4.5 &                0.72 \\
J0029.4$+$2051 &                 0.12 &                           5.6 & $7.2 \times 10^{6}$ & $5.4 \times 10^{5}$ &   1.6 &                  3.1 &  $1.2 \times 10^{2}$ &                     S0 &                                            3.3 &                0.55 \\
J0039.1$-$2219 &                0.015 &                           7.0 & $3.0 \times 10^{6}$ & $1.0 \times 10^{2}$ &   1.6 &                 0.18 &  $3.5 \times 10^{2}$ &                     S0 &                                            4.4 &                 1.0 \\
J0049.5$-$4150 &                 0.60 &                          11.0 & $3.3 \times 10^{6}$ & $8.3 \times 10^{5}$ &   1.4 &                  4.6 &                 53.0 &                     S0 &                                            4.3 &                 0.30 \\
J0054.7$-$2455 &                 0.11 &                           3.0 & $4.6 \times 10^{8}$ & $7.7 \times 10^{4}$ &   1.5 &                  4.3 &                 48.0 &                   Ell2 &                                            6.0 &                0.45 \\
\end{longtable}
}
\vspace{2em}

{\tiny \begin{longtable}{ccccccccc}
\caption{{\small Results for the synchrotron peak frequency $\nu_{sync}^{peak}$ and CD obtained from the broadband SED modelling, along with the energy budget results, are presented. These include the luminosity carried by the jet for the non-thermal low-energy and total radiative components, $L_{sync}$ and $L_{rad}$, the luminosities of the electrons $L_{kin}$ and the magnetic fields $L_B$, the total jet luminosity $L_{tot}$, and the ratio of the magnetic to electron energy density $U_B/U_e$.}}
\label{table:energy-budget}\\
\toprule
     4FGL Name & $\nu_{sync}^{peak}$ [Hz] &    CD & $L_{sync}$ [erg s$^{-1}$] & $L_{rad}$ [erg s$^{-1}$] & $L_B$ [erg s$^{-1}$] & $L_{kin}$ [erg s$^{-1}$] & $L_{tot}$ [erg s$^{-1}$] &            $U_B/U_e$ \\
\midrule
\endfirsthead
\caption[]{\textit{continued}} \\
\toprule
     4FGL Name & $\nu_{sync}^{peak}$ [Hz] &    CD & $L_{sync}$ [erg s$^{-1}$] & $L_{rad}$ [erg s$^{-1}$] & $L_B$ [erg s$^{-1}$] & $L_{kin}$ [erg s$^{-1}$] & $L_{tot}$ [erg s$^{-1}$] &            $U_B/U_e$ \\
\midrule
\endhead
\midrule
\multicolumn{9}{r}{{Continued on next page}} \\
\midrule
\endfoot

\bottomrule
\endlastfoot
J0003.1$-$5248 &     $2.5 \times 10^{17}$ &  0.69 &      $4.3 \times 10^{42}$ &     $6.9 \times 10^{42}$ & $2.3 \times 10^{42}$ &     $1.4 \times 10^{47}$ &     $1.4 \times 10^{47}$ &                0.033 \\
J0008.4$-$2339 &     $6.7 \times 10^{16}$ &  0.22 &      $4.3 \times 10^{40}$ &     $5.1 \times 10^{40}$ & $1.2 \times 10^{41}$ &     $5.8 \times 10^{44}$ &     $5.8 \times 10^{44}$ &                0.012 \\
J0013.9$-$1854 &     $9.5 \times 10^{17}$ &  0.16 &      $5.0 \times 10^{41}$ &     $6.0 \times 10^{41}$ & $6.0 \times 10^{41}$ &     $1.0 \times 10^{46}$ &     $1.0 \times 10^{46}$ &                0.035 \\
J0029.4$+$2051 &     $4.3 \times 10^{17}$ &  0.91 &      $2.9 \times 10^{43}$ &     $5.0 \times 10^{43}$ & $2.1 \times 10^{42}$ &     $1.7 \times 10^{44}$ &     $2.2 \times 10^{44}$ &                0.021 \\
J0039.1$-$2219 &     $1.2 \times 10^{18}$ &   1.2 &      $5.5 \times 10^{42}$ &     $1.4 \times 10^{43}$ & $3.5 \times 10^{40}$ &     $4.5 \times 10^{44}$ &     $4.6 \times 10^{44}$ & $1.4 \times 10^{-4}$ \\
J0049.5$-$4150 &     $7.3 \times 10^{17}$ &   0.80 &      $2.5 \times 10^{45}$ &     $4.1 \times 10^{45}$ & $5.3 \times 10^{43}$ &     $3.2 \times 10^{44}$ &     $4.5 \times 10^{45}$ &                 0.18 \\
J0054.7$-$2455 &     $1.0 \times 10^{16}$ &  0.42 &      $7.7 \times 10^{41}$ &     $1.1 \times 10^{42}$ & $1.9 \times 10^{42}$ &     $5.5 \times 10^{43}$ &     $5.8 \times 10^{43}$ &                0.069 \\
J0059.3$-$0152 &     $7.3 \times 10^{17}$ &   0.40 &      $2.1 \times 10^{44}$ &     $2.8 \times 10^{44}$ & $1.0 \times 10^{43}$ &     $1.0 \times 10^{44}$ &     $3.9 \times 10^{44}$ &                0.096 \\
\end{longtable}
}

\FloatBarrier

\section{FITS File Structure}
\label{appendix:fits_table}

This section summarizes the contents of the FITS file published in Zenodo (doi: \href{https://zenodo.org/records/15778400}{10.5281/zenodo.15778399}) and in \url{https://www.ucm.es/blazars/ehsp}. The HDUs contained in the FITS file are summarized in Table \ref{tab:hdus}. Table 1 contains the best-fit parameter values obtained from the broadband SED modelling performed for the 124 sources in the sample. Table 2 contains the resulting synchrotron peak and IC peak frequencies, Compton dominance, jet luminosities of each component, and the magnetic to kinetic energy density ratio for the 124 sources in the sample. Table 3 includes the expected detection significance with CTAO for all the sources, as well as their redshift. Table 4 contains for each source in the sample all the multi-wavelength data that has been used to reconstruct the broadband SED and to perform the SED modelling. The description of the contents included in these tables is given in Tables \ref{tab:hdu1}, \ref{tab:hdu2}, \ref{tab:hdu3} and \ref{tab:hdu4}.

\begin{table}[H]
    \centering
    \begin{tabular}{cccc}
    \toprule
         HDU Index & Type & Shape & Description \\
    \midrule
         1 & BinTableHDU & 124R x 15C & Best-fit parameters resulting from the SED modelling \\ 2 & BinTableHDU & 124R x 10C & Other derived physical parameters and energetic report \\  3  & BinTableHDU & 124R x 4C & CTAO expected detection significance \\ 4 & BinTableHDU & 124R x 6C & Multi-wavelength SED data used for the modelling \\ 
    \bottomrule
    \end{tabular}
    \caption{HDUs contained in the FITS file.}
    \label{tab:hdus}
\end{table}

\begin{table}[H]
    \centering
    \begin{tabular}{cccc}
    \toprule
         Column Name & Data Type & Units & Description \\
    \midrule
         `Source\_Name' & bytes16 & -  & Source name given in the 4FGL catalogue \\ `B'           & float64 & G & Magnetic field strength \\  `R' & float64 & cm  & Radius of the emitting region \\ `R\_H' & float64 & cm & Distance from the emitting region to the central black hole \\  `theta' &  float64 & deg   & Jet viewing angle \\  `BulkFactor'  & float64  &  -  &  Bulk Lorentz factor of the electrons in the jet \\  `gmin' & float64   & - & Minimum Lorentz factor of the electron population \\  `gmax' & float64 & - & Maximum Lorentz factor of the electron population \\  `N' &  float64 & 1/cm$^{3}$ & Electron density \\ `gamma\_break' & float64  & - & Break Lorentz factor  \\  `p1' & float64 & -  & Spectral slope of the lower energy electron population \\ `p2' & float64  & - & Spectral slope of the higher energy electron population \\  `Host\_Galaxy' & bytes17 &  - &  Best-fit host galaxy model \\ `T\_host' & float64  & K  & Temperature of the host galaxy \\ `chisq' & float64 & -     & Chi square value of the fitting \\
    \bottomrule
    \end{tabular}
    \caption{Description of table 1 (HDU 1).}
    \label{tab:hdu1}
\end{table}

\begin{table}[H]
    \centering
    \begin{tabular}{cccc}
    \toprule
         Column Name & Data Type & Units & Description \\
    \midrule
         `Source\_Name' & bytes16 & -  & Source name given in the 4FGL catalogue \\ `nu\_sync\_peak' & float64   & Hz    & Frequency of the synchrotron peak \\ `nu\_IC\_peak'  & float64   & Hz   & Frequency of the inverse Compton peak \\ `CD'  & float64  & -  & Compton dominance \\ `jet\_L\_Sync'  & float64   & erg/s & Jet luminosity due to the synchrotron component \\ `jet\_L\_rad'   & float64   & erg/s & Jet luminosity associated with radiative mechanisms \\ `jet\_L\_B'  & float64  & erg/s & Jet luminosity due to the magnetic field \\ `jet\_L\_kin' & float64   & erg/s & Jet luminosity due to the electrons \\ `jet\_L\_tot' & float64   & erg/s & Total jet luminosity \\ `UB/Ue' & float64   & -  & Ratio of magnetic to electron energy density \\
    \bottomrule
    \end{tabular}
    \caption{Description of table 2 (HDU 2).}
    \label{tab:hdu2}
\end{table}

\begin{table}[H]
    \centering
    \begin{tabular}{cccc}
    \toprule
         Column Name & Data Type & Units & Description \\
    \midrule
         `Source\_Name' & bytes16 & -  & Source name given in the 4FGL catalogue \\ `CTAO\_significance' & float64   & -     & CTAO expected detection significance \\ `Redshift'          & float64   & -     & Redshift of the source \\ `Redshift\_reference'& bytes10   & -     & Reference of the redshift estimate \\
    \bottomrule
    \end{tabular}
    \caption{Description of table 3 (HDU 3).}
    \label{tab:hdu3}
\end{table}

\begin{table}[H]
    \centering
    \begin{tabular}{cccc}
    \toprule
         Column Name & Data Type & Units & Description \\
    \midrule
         `Source\_Name' & bytes16 & -  & Source name given in the 4FGL catalogue \\ `log10nu'  & float32   & Hz          & log nu values used to reconstruct the broadband SED \\ `log10nuFnu'    & float32   & erg/(cm$^2$ s) & log nuFnu values used to reconstruct the broadband SED \\ `unc\_log10nuFnu'& float32   & erg/(cm$^2$ s) & Uncertainty in log10nuFnu \\ `UL'            & bool      & -           & Upper limits \\ `instrument'    & bytes37   & -           & Instrument that took the corresponding SED data  \\
    \bottomrule
    \end{tabular}
    \caption{Description of table 4 (HDU 4).}
    \label{tab:hdu4}
\end{table}

\end{appendix}

\end{document}